\documentclass[cup6b]{cupbook}
\usepackage{graphicx}
\usepackage{aas_macros}
\newcommand{\Lsun}{L_{\odot}}
\newcommand{\Msun}{M_{\odot}}

\newcommand{\rhalf}{r_{1/2}}

\newcommand{\Mhalf}{M_{1/2}}

\newcommand{\Vmax}{V_{\rm max}}
\newcommand{\rmax}{r_{\rm max}}
\newcommand{\kms}{{\rm km}\,{\rm s}^{-1}}
\newcommand{\sigmatot}{\sigma_{_{\rm{tot}}}}
\newcommand{\sigmastar}{\sigma_\star}
\newcommand{\sigmalos}{\sigma_{\rm los}}
\newcommand{\ave}[1]{\langle #1 \rangle}

\newcommand{\gtrsim}{\lower.7ex\hbox{$\;\stackrel{\textstyle>}
{\sim}\;$}}
\newcommand{\lesssim}{\lower.7ex\hbox{$\;\stackrel{\textstyle<}
{\sim}\;$}}

\begin{document}

\pagenumbering{roman}
%\maketitle
%\tableofcontents
%\cleardoublepage
\pagenumbering{arabic}

\chapter{Notes on the Missing Satellites Problem \\  {\small James. S. Bullock (UC Irvine)}}

\abstract
 
 The Missing Satellites Problem (MSP)  broadly refers to the overabundance of predicted Cold Dark Matter (CDM) subhalos compared to satellite galaxies known to exist in the Local Group.   
 The most popular interpretation of the MSP is that the smallest dark matter halos in the universe are extremely inefficient at forming stars.  The question from that standpoint is to identify the feedback source that makes small halos dark and to identify any obvious mass scale where the truncation in the efficiency of galaxy formation occurs.

Among the most exciting developments in near-field cosmology in recent years is the discovery of a new population satellite galaxies orbiting the Milky Way and M31.  Wide field, resolved star surveys have more than doubled the dwarf satellite count in less than a decade, revealing a population of ultrafaint galaxies that are less luminous that some star clusters.    For the first time, there are {\em empirical} reasons to believe that there really are more than 100 missing satellite galaxies in the Local Group, lurking just beyond our ability to detect them, or simply inhabiting a region of the sky that has yet to have been surveyed.  

Remarkably, both kinematic studies and completeness-correction studies seem to point to a characteristic potential well depth for satellite subhalos 
that is quite close to the mass scale where photoionization
and atomic cooling should limit galaxy formation.  Among the more pressing problems associated with this interpretation is to understand the selection biases that limit our ability to detect the lowest mass galaxies.  
The least massive satellite halos are likely to host stealth galaxies with very-low surface brightness and this may be an important limitation in the hunt for low-mass fossils from the epoch of reionization.

\section{Perspective and Historical Context}

These lecture notes focus on a potential problem for the standard model of structure formation, but before delving into this issue in depth, it is worth providing a bit of perspective.  Cosmological observations ranging from the cosmic microwave background (Komatsu et al. 2010) to large-scale galaxy clustering (Reid et al. 2010) to the Lyman-$\alpha$ forest (Viel et al. 2008) all point to a concordance cosmological model built around Dark Energy + Cold Dark Matter ($\Lambda$CDM).    Specifically, direct constraints on the power spectrum of density fluctuations suggest that structure formation in the universe is hierarchical (Press \& Schechter 1974; White \& Rees 1978), at least down to the scale of large galaxies.   This is a remarkable fact and, for those of us interested in galaxy formation, it provides some confidence that we are not wasting our time attempting to build galaxies within the scaffolding  provided by a CDM-based cosmology (Blumenthal et al.  1984; Davis et al. 1985).     On still smaller scales, stellar streams and clumps appear to fill the stellar halos of the Milky Way and M31 (Ivezi{\'c} et al. 2000; Newberg et al. 2002; Ferguson et al. 2002; Majewski et al. 2003; Zucker et al. 2004; Guhathakurta et al. 2006; Ibata et al. 2007; Bell et al. 2008; Watkins et al. 2009; McConnachie et al. 2009) in accordance with expectations from $\Lambda$CDM-based models of dwarf galaxy accretion and disruption (Bullock, Kravtsov \& Weinberg 2001; Bullock \& Johnston 2005; Cooper et al. 2010).  Even if the underlying makeup of the universe is somehow different than this, it must look very much like $\Lambda$CDM from the scale of the horizon down to $\sim 50$ kpc or so.  It is within this context that we consider the Missing Satellites Problem in CDM.

A defining characteristic of CDM-based hierarchical structure formation is that dark matter halos are assembled via mergers with smaller systems.  In the standard framework of CDM, density inhomogeneities in the early universe are small compared to unity and characterized by a Gaussian distribution with variance $\sigma$.  If we consider density fluctuations within spheres of radius $R$ and associated mass $M \propto R^3$, we can characterize the linear fluctuation amplitude as
\begin{equation}
\sigma(M,a) \propto  D(a) \, F(M) \,
\end{equation}
where $a = (1+z)^{-1}$ is the expansion factor and $D(a)$ quantifies the 
gravitational amplification of fluctuations, which grow as $D(a) \propto a$ at early times when the universe is near critical density.
CDM predicts  that $F(M)$ bends as $F \propto M^{-\alpha}$ with  $\alpha = 2/3$ on large scales to $\alpha \rightarrow 0$ (logarithmically) on the smallest mass scales (Blumenthal et al. 1984).  
At any given redshift, the typical mass-scale that is collapsing, $M_*(a)$, 
can be estimated by determining the non-linear scale via $\sigma(M_*,a)  \sim 1$, which implies $M_*(z) \sim D(a)^{1/\alpha}$. 
This means low-mass scale fluctuations eventually break off from the expansion first.   More massive objects are assembled later (at least in part) by many smaller-scale pre-collasped regions (Press \& Schechter 1974).  The normalization of the power spectrum is often quantified by $\sigma_8 =
\sigma(M = M_8,a = 1)$, whre $M_8$ is the mass enclosed within a sphere of radius $8 h^{-1}$ Mpc.  Structure of a given mass tends to collapse earlier for a higher value of $\sigma_8$.

One implication of this hierarchical scenario is that 
small dark matter halos
collapse at high  redshift, when the universe  is very dense, and as a result have high density concentrations.  When these  halos
merge into larger hosts, their high core densities allow them to resist the
strong tidal  forces that  would otherwise act to  destroy them.   While gravitational
interactions  act to strip mass from merged progenitors, 
a significant fraction of these small halos survive as bound substructures (see, e.g. Zentner \& Bullock 2003).  

Kauffmann, White, \& Guiderdoni (1993) were the first to show using a self-consistent treatment that substructure would likely survive the merging process with abundance in CDM halos.  They used a semi-analytic model to show that subhalos within cluster-size dark matter halos will have approximately the correct number and mass spectrum to account for cluster {\em galaxies}.  On the other hand, the same calculation demonstrated that Milky-Way size halos should host a large number of satellite subhalos, with over $\sim 100$ objects potentially massive enough to host observable satellite galaxies (with $L > 10^6 \Lsun$), even when allowing for fairly substantial supernovae feedback.   Given that there are only $\sim 10$ satellites brighter than this around the Milky Way, the authors concluded that dwarf galaxy formation would need to be suppressed by some large factor in order to explain the discrepancy.  Nevertheless, this dwarf satellite issue did not gain significant attention in the community until these initial semi-analytic estimates were (more or less) verified by direct numerical calculations some $\sim 5$ years later (Klypin et al. 1999a; Moore et al. 1999).  It was only then that the Missing Satellites Problem (MSP) gained urgency.

Implicit in the title Missing Satellites  is the notion of prediction   -- the satellites are expected by theory but are not seen.  It is not surprising then that the problem was named by theorists.~\footnote{ 
A strict empiricist  may have called it ``The Overabundant Substructure Problem", but this is not what the MSP is now called
even though many may interpret it this way.}
Kylpin et al. (1999a) titled their seminal paper
on the topic ``Where are the Missing Galactic Satellites?".  The notion is that CDM is a successful 
theory and that the  substructure prediction is robust and to be taken seriously.  A similar 
sentiment was echoed in the nearly coeval paper 
by Moore et al. (1999): This is a {\rm robust} prediction of CDM, so why don't we see more dwarfs?  
It is indeed the robustness of the Klypin et al. and Moore et al. calculations that demanded attention to the problem that was sketched by Kauffmann et al. (1993). 
More recent simulations, now with a factor of $\sim 10^3$ more particles, have effectively confirmed these first calculations (Diemand et al. 2008; Springel et al. 2008).  The mass function of substructure is predicted to rise steeply to the smallest masses, while the luminosity function of observed dwarf satellites is fairly flat.
In this set of lecture notes,  I will touch on some of the physical mechanisms that may act to suppress galaxy formation within the smallest halos, but will 
focus primarily on the best way compare predictions with observations.  Specifically, the zeroth order concern for any $\Lambda$CDM (or variant) model that wishes to match the Local Group satellite population is to correctly reproduce the mass-luminosity relationship obeyed by  faint galaxies.  This necessarily relies on mass determinations.   Once we have robust mass determinations for the dwarfs, we can begin to test specific models for their formation and evolution.
Section 1.4 provides an overview of recent efforts to constrain the mass-luminosity relation for dwarf galaxies.

The most important observational development in MSP studies in the last decade has been the discovery of a new population of faint satellite galaxies.
Approximately twenty-five new dwarf galaxy companions of the Milky Way and M31
have been discovered since 2004, more than doubling the known satellite 
population in the Local Group  in five years (Willman et al. 2005; Zucker et al. 2006; Grillmair 2006, 2009; Majewski et al. 2007; Belokurov et al. 2007, 2009; Martin et al. 2009; see Willman 2010 for a recent review).   The majority of these newly-discovered 
dwarfs are less luminous than any galaxy previously known.  The most extreme of these, the ultrafaint Milky Way dwarfs,  
have luminosities~\footnote{We will assume V-band luminosities throughout} smaller than an average globular cluster $L \simeq 10^2 - 10^4$ L$_\odot$, and were discovered
by searches for stellar overdensities in the wide-field maps of the Sloan Digital Sky Survey (SDSS) and the
Sloan Extension for Galactic Understanding and Exploration (SEGUE).
Follow-up kinematic observations showed that these tiny galaxies have surprisingly high stellar velocity dispersions 
for their luminosities and sizes ($\sigma_\star \sim 5 \, \kms$; Martin et al. 2007; Simon \& Geha 2007; Simon et al. 2010)
and subsequent mass modeling has shown that
they are the most dark matter dominated galaxies known (Wolf et al. 2010; Martinez et al. 2010).
Remarkably, these extreme systems are not only the
faintest, most dark matter dominated
galaxies in the universe but they are also the most metal poor  stellar systems yet studied (Kirby et al. 2008; Geha et al. 2009).
All of the new dwarfs, with the exception of Leo T at a distance of more than 400 kpc, have negligible atomic hydrogen fractions (Grcevich \& Putman 2009).  
Perhaps the most exciting aspect of these recent discoveries is that completeness corrections point to a much
larger population of undiscovered dwarfs (Koposov et al. 2008; Tollerud et al. 2008; Walsh et al. 2009; Bullock et al. 2010).    For the first time
we have empirical evidence to suggest that there really are missing satellite galaxies in halo of the Milky Way and this is the subject of Section 1.5.   These missing objects are likely extremely dark matter dominated, but too dim and/or too diffuse to have been discovered (yet).

\section{Counting and Defining Dark Matter Halos in CDM}

\subsection{Galaxy Halos}

The standard paradigm of galaxy formation today posits that all (or at least an overwhelming majority) of galaxies reside near the centers of
extended dark matter halos.  The outer edges of the halos are somewhat arbitrarily defined, but a typical convention is that 
halos {\em in the field}
are characterized by a radius $R_{\rm h}$ and mass $M_{\rm h}$ that obey 
$M_{\rm h} = \Delta_{\rm h} \, \rho_m \, (4 \pi \,  / 3) \, R_{\rm h}^3$,   
such that average density  within the halo exceeds the background $\rho_m$ by a factor $\Delta_{\rm h} \sim 200$, which is motivated by the virialized overdensity obtained in approximate spherical collapse models (see, e.g. Bryan \& Norman 1998). ~\footnote{For our purposes the precise value of
$\Delta_{\rm h}$ does not matter, but in principle one needs to be careful about the definition when doing comparisons, especially when considering halos more massive than $M_{\rm h} \sim M_*$.}
With mass defined in this way, N-body simulations reveal that the halo mass function is remarkably well characterized by
(see Tinker et al. 2008 and references therein):
\begin{equation}
\frac{ { \rm d} \, n}{ {\rm d} M_{\rm h}} = f_{\rm h}(\sigma) \, \frac{\rho_m}{M_{\rm h}^{2}} \, \frac{d \ln \sigma^{-1}}{d \ln M_{\rm h}} ,
\end{equation}
where $f_{\rm h}(\sigma)$ is effectively a fitting function that approaches a constant at low mass ($M_{\rm d} \ll M_*$ or $\sigma \gg 1$) and becomes exponentially suppressed for halos that are too large to have collapsed ($M_{\rm d} \gg M_*$ or $\sigma \ll 1$).\footnote{Note that the {\em linear} mass variable $M$ in $\sigma(M)$ has somewhat magically replaced by the 
virial mass variable $M \rightarrow M_{\rm h}$ in this equation.  It is not outrageous to think that the two should be related, but an exact correspondence would
be bizarre given the simplicity of spherical collapse.   The function $f_{\rm h}(\sigma)$ can be regarded as a fudge factor that makes up for this difference.}  
One important observation (which has been recognized at least since White \& Rees 1978) is that the mass function rises steadily towards smaller masses $dn/dM \sim M^{-2}$, while the luminosity function or baryonic mass function of galaxies in the real universe does not rise as steeply.   Clearly a one-to-one
correspondence between mass and luminosity must break down at small masses.

\begin{figure*}[!t!]
\centering
\begin{tabular}{ccc}
\includegraphics[angle=-90,width=0.9\linewidth]{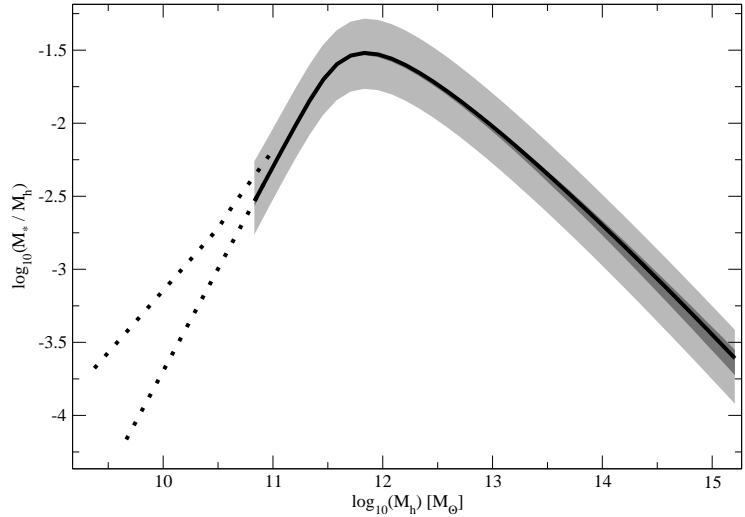} \hspace{1in}
\end{tabular}
\caption{Relationship between stellar mass $M_*$ and halo mass $M_{\rm h}$ as quantified by the abundance matching analysis of Behroozi et al. (2010),
which includes uncertainties associated with observed stellar mass functions and possible scatter in the relationship between halo mass and stellar mass (shaded band).  The lower dashed line is the extrapolated relationship that results from assuming the stellar mass function continues as a power law for $M_* < 10^{8.5} \Msun$, using data from the full SDSS (Li \& White 2009).  The upper dashed line is extrapolated under the assumption that the slope of the stellar mass function has an upturn at low masses to $\alpha = -1.8$, as found by Baldry, Glazebrook, \& Driver (2008).  }
\label{f:am}
\end{figure*}

A more precise way of characterizing the relationship between galaxies and their halos comes from the technique known as abundance matching.
One simply asks the question: what mass-luminosity relationship must I impose in order to reproduce the observed luminosity function of galaxies?
Remarkably, two-point clustering statistics of bright galaxies 
can be explained fairly well under the simple assumption that galaxy luminosity $L$ or stellar mass $M_*$ 
maps to dark matter halo mass
 $M_{\rm h}$ in a nearly monotonic way (see Kravtsov et al. 2004; Conroy \& Wechsler 2009; Moster et al. 2010; Behroozi et al. 2010 and references therein).  Figure \ref{f:am} illustrates the required mapping by showing the ratio
 of galaxy stellar mass to halo virial mass ($M_*/M_{\rm h}$) as a function of halo mass (P. Behroozi, 2010, private communication).  Dark gray shading indicates statistical and sample variance errors and the light gray shading includes systematic errors.   The shaded band is truncated at small stellar masses where the mass functions become incomplete in the Sloan Digital Sky Survey ($M_* \simeq 10^{8.5} \Msun$).  It is clear from this figure that the efficiency of galaxy formation must peak in halos with masses $\sim 10^{12} \Msun$, 
and to become increasingly inefficient towards smaller and larger halo masses.    An understanding of this behavior 
-- how and why it happens -- remains a fundamental goal in galaxy formation.

The two dashed lines in Figure 1 bracket reasonable low-mass extrapolations of the relationship defined at larger masses, with the upper line defined by a power-law extrapolation of the observed stellar mass function from Li \& White (2009) and the lower dashed corresponding to a case where the stellar mass function has
an upturn, as advocated by Baldry et al. (2008).  At small masses, the implied range of scalings is $M_* \propto M_{h}^{p}$ with $p \simeq 2.5 - 1.8$.
For example, at the edge of the figure, the power-law extrapolation suggests that $M_{\rm h} = 10^9 \Msun$ halos should host galaxies 
with stellar masses between $M_* \simeq 5 \times 10^{4} \Msun$ and $M_* \simeq  3 \times 10^5 \Msun$.  As we shall see, halo masses of $M_{\rm h} \sim 10^9 M_\odot$ ($\Vmax \simeq 30 \kms$) are approximately those required to match the kinematics of $L \sim 10^5 L_\odot$  dwarf galaxies (Strigari et al. 2008).  Of course, there is no real physics in this extrapolation, but it is encouraging that
when one imposes this simple, empirically motivated scaling, we get numbers that are not far off from those required to match dwarf satellite
observations  (Busha et al. 2010; Kravtsov 2010).  A more general comparison of abundance-matching masses and kinematically-derived masses
is presented in Tollerud et al. (2010).

\subsection{Subhalos}

Galaxies that are as faint as $L \sim 10^5 L_\odot$ are very difficult to detect, and, as we discuss below (see Figure \ref{f:Rcomp}),  we are only reasonably confident in our census  of these objects to distances comparable to the expected virial radius of the Milky Way's halo, and within small, deeply surveyed regions within the virial radius of M31 (Majewski et al. 2007; Ibata et al. 2007; McConnachie et al. 2008; Martin et al. 2009).  In addition, these galaxies appear to be embedded within dark matter halos (Mateo 1998; Simon \& Geha 2007; Walker et al. 2009; Wolf et al. 2010; Kalirai et al. 2010; Collins et al. 2010).  The implication is that in order to confront the faintest galaxies theoretically in the context of $\Lambda$CDM, we must discuss them as embedded within subhalos -- dark matter halos that are bound and within $R_{\rm h}$ of a larger host halo.

Subhalos (unlike field halos) are not naturally characterized by a virial mass, $M_{\rm h}$, because they tend to be truncated by tidal forces at radii smaller than their over-density-defined virial radii (e.g., Klypin et al. 1999b; Kazantzidis et al. 2004; Pe{\~n}arrubia et al. 2008).  The most common way one characterizes a subhalo is to use $V_{\rm max}$, the maximum value of the circular velocity curve $V_c(r) = (G \, M(r) \, / r)^{1/2}$, which itself is measured in spheres, stepping out from the subhalo center of mass.  The radius $\rmax$ where $\Vmax$ occurs is typically set by tidal stripping, which more readily unbinds material from the outer parts of galaxy halos than the inner parts (Bullock \& Johnston 2007; Kazantzidis et al. 2010).
The choice of $V_{\rm max}$ as a quantifier is particularly nice because it is fairly robust to the details of the halo finding algorithm used.  Another choice is to assign subhalos a mass directly.  Unfortunately, {\em total} mass assignment to subhalos is somewhat subjective and can result in differences from halo finder to halo finder, but reasonable choices involve 1) defining the subhalo boundary at the radius where its density profile approximately reaches the mean background density of the host halo; 2)  estimating a local tidal radius iteratively;  or 3) removing unbound particles from the subhalo iteratively in such a way that a boundary emerges fairly naturally (see, e.g. Springel et al. 2008 Figure 14 and related discussion).

\begin{figure*}[t!]
\centering
\begin{tabular}{ccc}
\includegraphics[width=4in]{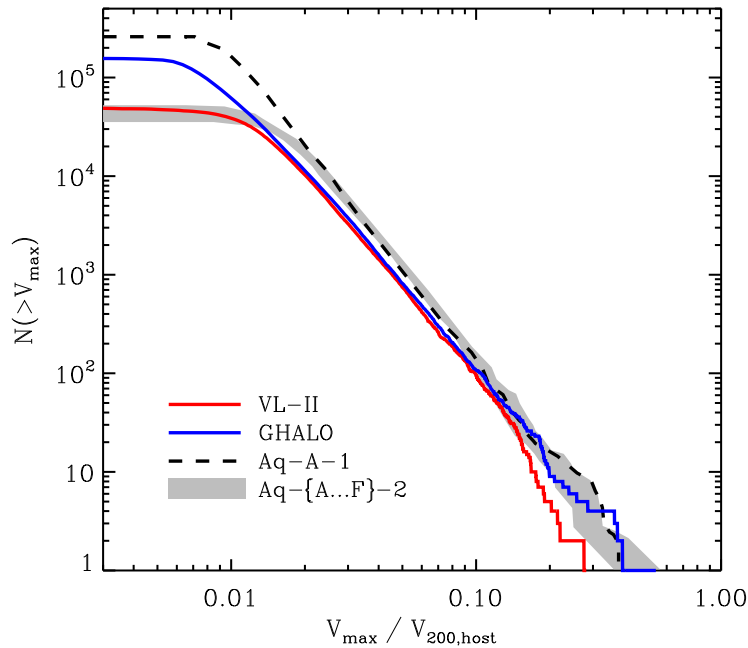}
\end{tabular}
\caption{Cumulative subhalo $\Vmax$ function within $R_{h} \simeq 400$ kpc for the three highest resolution simulations of Milky-Way size halos from
Kuhlen (2010, private communication), shown here relative to the circular velocity of each host at its outer radius $V_{\rm h} \simeq 140 \, \kms$.    The solid red line shows VL II subhalos from Diemand et al. (2008), the dashed black line shows Aq-A subhalos from Springel et al. (2008) and the solid blue line shows GHALO subhalos from Stadel et al. (2009).  The shaded band provides an estimate of halo-to-halo scatter from a series of lower resolution halos (Aq suite, Springel et al. 2008).  The small normalization difference between the GHALO/VL II and Aq halos is likely a result of their having different power spectrum normalizations (Zentner \& Bullock 2003). The most recent determination of the normalization WMAP-7 (Komatsu et al. 2010) is intermediate between the adopted normalizations of the simulations shown here.}
\label{f:Vmaxfunc}
\end{figure*}

Figure \ref{f:Vmaxfunc} summarizes cumulative subhalo $V_{\rm max}$ functions from the three highest resolution $\Lambda$CDM simulations of Milky-Way size halos that have yet been run
(VL II -- Diemand et al. 2008; Aq-A -- Springel et al. 2008; and GHALO -- Stadel et al. 2009; Kuhlen et al. 2010).  Each simulation contains $N_p \simeq 10^9$ particles within the halo radius $R_{\rm h} = R_{200}$\footnote{defined via $\Delta = 200$ over the {\em background} density} and the subhalos plotted are those within $R_{\rm h}$. Because the three halos have slightly different masses ($M_{\rm h}  = 1.3$, $1.9$, and $2.5  \times 10^{12} \Msun$,  for GHALO, VL II, and Aq-A, respectively)   the $\Vmax$ values in Figure \ref{f:Vmaxfunc} have been scaled to $V_{\rm h} = V_{200} = (G M_{\rm h}/R_{h})^{1/2}$ for each system, where
$R_{\rm h} = 347$,  402, 433 kpc and $V_{\rm h} \simeq 127$, $142$, and $157 \, \kms$, for the three hosts.    The shaded band summarizes the results of a series of lower resolution simulations ($N_p \simeq 2\times 10^8$) presented in Springel et al. (2008).  The flattening seen in the cumulative $V_{\rm max}$ functions are entirely a result of resolution limits, as can be seen by comparing the shaded band of lower resolution Aq halos to the dashed line, which is the highest resolution Aq halo.     

The most striking feature of the comparison shown in Figure \ref{f:Vmaxfunc} is that the simulation lines basically agree.  The small normalization shift towards higher number counts for the Aq halos is likely due to the fact that 
the Aq runs use a higher normalization ($\sigma_8 = 0.9$) than the VL-II and GHALO runs ($\sigma_8 = 0.74$).  The difference seen is in line with the analytic expectations of Zentner \& Bullock (2003), who explored how changes to the power spectrum should affect substructure counts. 
Note that the most recent WMAP-7 preferred  value for the normalization is $\sigma_8 = 0.81$, which is 
intermediate between the two sets of simulations shown here.  Overall, the $V_{\rm max}$ functions are well-fit by a power law: 
\begin{equation}
N(>V_{\rm max}) \simeq 0.15 \left( \frac{V_{\rm max}}{V_{\rm h}}\right)^{-2.94} \, , 
\label{eq:nofv}
\end{equation}
where the normalization has been chosen to be intermediate between the two sets of simulations shown.  
For a Milky-Way size host ($V_{\rm h} \simeq 140 \, \kms$; Xue et al. 2008; Gnedin et al. 2010) we expect $N \simeq 350$ subhalos with $V_{\rm max} > 10 \kms$ within $R_{\rm d} \simeq 400$ kpc of the Milky Way.  This number matches well with the
predictions from the first MSP papers more than a decade ago (Klypin et al. 1999; Moore et al. 1999).  What we also see is that the count is expected to rise for smaller halos,  and there is no sign of a physical break.  The highest resolution simulation shown in Figure \ref{f:Vmaxfunc} has already resolved more than $10^5$ subhalos, and there is no reason to expect the numbers to stop rising in the absence of another scale in the problem.

\begin{figure*}[t!]
\centering
\begin{tabular}{ccc}
\includegraphics[width=4in]{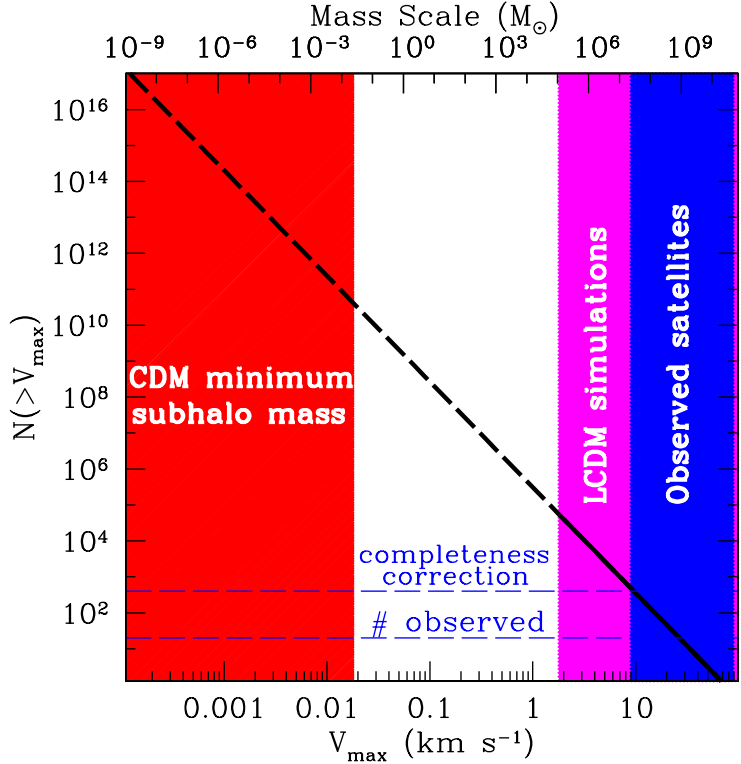}
\end{tabular}
\caption{Illustrative sketch of the expected cumulative CDM substructure abundance within the Milky Way's halo. The line is Equation \ref{eq:nofv} for a host halo of circular velocity $V_{\rm h} = 140 \, \kms$ at $R_{\rm h} = 400$ kpc (for which we would expect a maximum circular velocity of about 200 $\kms$).  
  The line becomes dashed where we are extrapolating the power-law beyond the resolving power of state-of-the-art numerical simulations.  The vertical red band provides an indication of where we expect this power-law to break for popular CDM dark matter candidates.   The lower horizontal dashed line shows the
number of Milky Way satellite galaxies known while the upper dashed line is an estimate of the total number of satellite galaxies that exist within 400 kpc of the Milky Way corrected for luminosity-bias and sky coverage limitations of current surveys (Tollerud et al. 2008).  The fact that the upper horizontal line intersects the edge of the vertical blue band at about the location of the CDM prediction is quite encouraging for the theory.}
\label{f:Extrap}
\end{figure*}

The only other obvious scale in CDM is the filtering mass in the power spectrum, where fluctuations are suppressed on the scale of the comoving horizon at the time of kinetic decoupling of the dark matter in the early universe (Loeb \& Zaldarriaga 2005; Bertschinger 2006):
\begin{equation}
M_{\rm min} = 10^{-4} \left( \frac{T_d}{10 MeV}\right)^{-3} \, \Msun .
\end{equation}
The kinetic decoupling temperature $T_d$ depends on the scattering interactions of dark matter with standard model fermions, and is therefore sensitive to the microphysical dark matter model itself (Profumo et al. 2006).  Martinez et al. (2009) showed that the popular Constrained Minimal Supersymmetric Standard Model (CMSSM) with neutralino dark matter allows $M_{\rm min} \simeq 10^{-9} - 10^{-6} \, \Msun$, after accounting for the entire presently viable parameter space of the CMSSM, including low-energy observables, the relic abundance ($\rho_m$), and direct constraints on standard model quantities. The implication is that there is good reason to suspect that the mass spectrum of substructure in the Milky Way halo continues to rise down to the scale of an Earth mass, and even smaller.

Figure \ref{f:Extrap} provides a cartoon-style estimate of the total substructure abundance within the Milky Way's halo down to the scale of the minimum mass.  Plotted is  the $N(>V_{\rm max})$ relation from Equation \ref{eq:nofv}
for a Milky-Way size host.  The line becomes dashed where we are extrapolating the power-law beyond the resolving power of state-of-the-art numerical simulations.  The upper axis provides a mass scale using bound-mass to $\Vmax$ relation for subhalos found by Springel et al. (2008):  $M_{\rm sub} = 3.4 \times 10^7 \Msun (\Vmax/10 \, \kms)^{3.5}$.  
The vertical red band provides an indication of where we expect this power-law to break.
Here, I have taken the $M_{\rm min}$ range from Martinez et al. (2009) and expanded it slightly to allow for uncertainties associated with going form the linear power spectrum to subhalos, and to qualitatively allow for more freedom in the dark matter model (e.g. Profumo et al. 2006). 
For comparison, the magenta band shows the mass (or velocity) scale that has been directly resolved in simulations and the vertical blue band on the far right indicates current range of $\Vmax$ values that are consistent with kinematic constraints from observed Milky Way satellite galaxies (as discussed below and in Pe{\~n}arrubia et al. 2008).

Based on this extrapolated power-law and the range of minimum subhalo masses,  we  expect a total of $\sim 10^{11}$ to $10^{17}$ bound subhalos to exist within $400$ kpc of the Sun!  Compared to the total number of satellite galaxies known ($\sim 20$; lower dashed line) or even an observational completeness corrected value ($\sim 400$; upper dashed line from Tollerud et al. 2008), we are very far from discovering all of the subhalos that are expected in CDM.  

The most extreme way one might characterize the MSP is to say that CDM predicts $> 10^{11}$ subhalos while we observe only $\sim 10^2$ satellite galaxies.
While this statement is true, it should not worry anyone because the vast majority of those CDM subhalos are less massive than the Sun.
No one expects a satellite galaxy to exist within a dark matter halo that is less massive than a single star.     We  {\em know} that galaxy formation must  truncate in dark matter structures smaller than some mass.~\footnote{A conservative requirement would be $M_{\rm h} > (1/f_b)  \Msun \sim 5 \Msun$ 
in order to host a ``galaxy" containing one solar mass star}
The question is what is that mass?  Is there a sharp threshold mass, below which galaxy formation ceases?  Can we use  the Local Group satellite population as a laboratory for galaxy formation on small scales?  

Physical processes like supernova feedback ($\Vmax \sim 100 \, \kms$), heating from photoionization ($\Vmax \sim 30 \, \kms$), or the ability for gas to cool ($\Vmax \sim 15 \, \kms$ for Lyman-alpha and $\sim 5 \, \kms$ for H$_2$) each imposes a different mass scale of relevance (Dekel \& Silk 1986; Efstathiou 1992; Tegmark et al. 1997; Bullock et al. 2000; Ricotti et al. 2001).     If, for example, we found evidence for very low-mass dwarf galaxies $\Vmax \sim 5 \, \kms$ then these would be excellent candidates for primordial $H_2$ cooling `fossils' of reionization in the halo  (Madau et al. 2008; Ricotti 2010).   We would like to measure the mass-scale of the satellites themselves and determine if the number counts of observed satellites are consistent with what is expected given their masses.  Unfortunately, it turns out that this comparison is not as straightforward as one might naively expect.

\section{Defining the Problem}

\subsection{Maximum Circular Velocity $\Vmax$}

From the perspective of numerical simulations, a  $\Vmax$ function such as that shown in Figure \ref{f:Vmaxfunc} provides a natural and robust way to characterize  the substructure content of a dark matter halo.  Ideally, we would like to put the satellite galaxies of the Milky Way and M31 on this plot in order to make a direct comparison between theory and data.  Unfortunately, $\Vmax$ is not directly observable for the dwarf satellites. Instead, the stellar velocity dispersion, $\sigmastar$, is the most commonly-derived kinematic tracer for dwarf spheroidal (dSph) satellite galaxies in the Local Group.~\footnote{The best studied dwarfs have individual velocity measurements for more than a thousand stars (Walker et al. 2009) and this allows for more nuanced studies of their kinematics involving velocity dispersion profiles, higher-order moments of the velocity distribution, and explorations of the underlying distribution function shape.}  dSph velocity dispersion profiles tend to be flat with radius (Walker et al. 2009) but for concreteness, I will use $\sigmastar$ to denote the luminosity-weighted line-of-sight velocity dispersion of the system.  

The first MSP papers estimated $\Vmax$ values for the dwarfs by assuming that their stellar velocity dispersion was equal to the dark matter velocity dispersion within the central regions of their host halos: $\sigmastar = \sigma_{\rm dm}^{{\rm SIS}} = \Vmax/\sqrt{2}$ (Moore et al. 1999) and $\sigmastar = \sigma_{\rm dm}^{{\rm NFW}} \simeq \Vmax/\sqrt{3}$ (Klypin et al. 1999).  This was a reasonable start, especially given that the main point was to emphasize an order-of-magnitude discrepancy in the overall count of objects.  Nevertheless, for detailed comparisons it is important to realize $\sigmastar \ne \sigma_{\rm dm}$.~\footnote{This simple fact has caused a lot of confusion in the literature, especially when it relates to attempts to ``measure" phase space densities of dark matter in dwarfs, 
which cannot be done without appealing to theoretical expectations for how and how far the dark matter halos extend beyond the stellar extent of the dwarfs.}  Specifically, since the stars are bound to the {\em central} regions of dark matter halos, while the dark matter particles can orbit to much larger radii, we generally expect $\sigmastar \lesssim \sigma_{\rm dm}$.  This fact motivated the idea that the dwarf galaxies could be more massive than originally supposed, and perhaps populate only the most massive subhalos of the Milky Way (Stoehr et al. 2002; Hayashi et al. 2003; and Kravtsov et al. 2004b, who discussed a more nuanced model associated with mass prior to accretion).
Below I demonstrate explicitly that the only relationship that can be made without imposing a theoretical model for subhalo structure is $\Vmax \ge \sqrt{3} \sigmastar$ (Wolf et al. 2010).  Cases where $\Vmax \simeq \sqrt{3} \sigmastar$ correspond to those where the edge of the stellar distribution is very close to $\rmax$.  Because $\rmax$ is typically set by tidal truncation in CDM halos, this last equivalence is expected to hold in cases where 
the stars have be affected or are beginning to be affected by tidal stripping/stirring in the Milky Way (as shown numerically by Kazantzidis et al. 2010).

One of the problems with placing dSph data on a $\Vmax$ function plot is that the mapping between $\sigmastar$ and $\Vmax$ depends sensitively on what one assumes about the  density profile structure of the subhalos.  The stellar velocity dispersion data probe only the potential within the stellar extent, and one must extrapolate beyond that point in order to estimate $\Vmax$.  
For the current population of well-studied Milky Way dSph galaxies (see Table 1 in Wolf et al. 2010), the median (3d) half-light radius is $\rhalf \simeq 300$ pc,
while rotation curves of $\Lambda$CDM subhalos with $\Vmax > 15 \, \kms$ can peak at radii ranging from $\sim 1000 - 6000$ pc depending on the value of $\sigma_8$ (Zentner \& Bullock 2003; Figure 21).  For a typical dwarf, this  amounts to a factor of $\sim 3$ to $\sim 20$ extrapolation in the assumed mass profile from the point where it is constrained by data (and it can be much worse for the smallest dwarfs).   If we restrict ourselves to $\Lambda$CDM models with $\sigma_8 = 0.8$ we expect that {\em median} rotation curves will peak at  radii ~\footnote{Estimated from the results of Springel et al. (2008) and Diemand et al. (2008) by chosing an intermediate normalization}
\begin{equation}
 r_{\rm max} \simeq 1100 \, {\rm pc} \, \left( \frac{V_{\rm max}}{15 \, \kms}\right)^{1.35} \, .
 \label{eq:vm}
 \end{equation}
Even with fixed $\sigma_8$, the extrapolations are large and sensitive to halo-to-halo scatter in profile parameters (the $68 \%$ scatter is about a factor of  $\sim 2$ in $\rmax$ at fixed $\Vmax$ according to Springel et al. 2008).  For the most extreme cases of ultra-faint dwarf galaxies, the stellar extent is so small ($\rhalf \sim 30-40$ pc) that we must rely on a factor of $\sim 35$ extrapolation in $r$ in order to reach a typical $\Vmax$ value at $\rmax \simeq 1100$ pc.   One begins to wonder whether there might be a better option than $\Vmax$ for comparing theory to data.

\subsection{Kinematic Mass Determinations}

What quantity is best constrained by stellar velocity dispersion data?  Consider
a spherically-symmetric galaxy that is in equilibrium 
with stellar density distribution $\rho_*(r)$ and radial velocity profile $\sigma_r(r)$ that is
embedded within total mass profile $M(r)$.  The Jeans equation is conveniently written as
\begin{equation}
M(r) = \frac{r \: \sigma_r^2}{G} \left (\gamma_\star+\gamma_\sigma - 2\beta \right ),
\label{eq:massjeans}
\end{equation}
where $\beta(r) \equiv 1- \sigma_t^2 / \sigma_r^2$ characterizes the tangential velocity dispersion and 
$\gamma_{\star} \equiv - {\rm d}\ln \rho_\star / {\rm d} \ln r$ and $\gamma_{\sigma} \equiv - {\rm d}\ln \sigma_r^2 / {\rm d} \ln r$. 
In principle, $\sigma_r(r)$ can be inferred from the observed line-of-sight velocity dispersion profile $\sigma_{\rm los}(R)$, but this geometric deconvolution depends on the unknown function $\beta(r)$.  At this point one may begin to get worried:
 uncertainties in $\beta$ will affect both the mapping from observed $\sigmalos$ to $\sigma_r$ {\em and} the derived relationship between $M(r)$ and $\sigma_r$  in Equation \ref{eq:massjeans}.  Given that the dSph galaxies are well outside the regime where we can legitimately assume that mass-follows light, are we forced to assume that $\beta = 0$ in order to derive any meaningful mass constraint?  Thankfully, no.

\begin{figure*}
\includegraphics[width=0.48\linewidth]{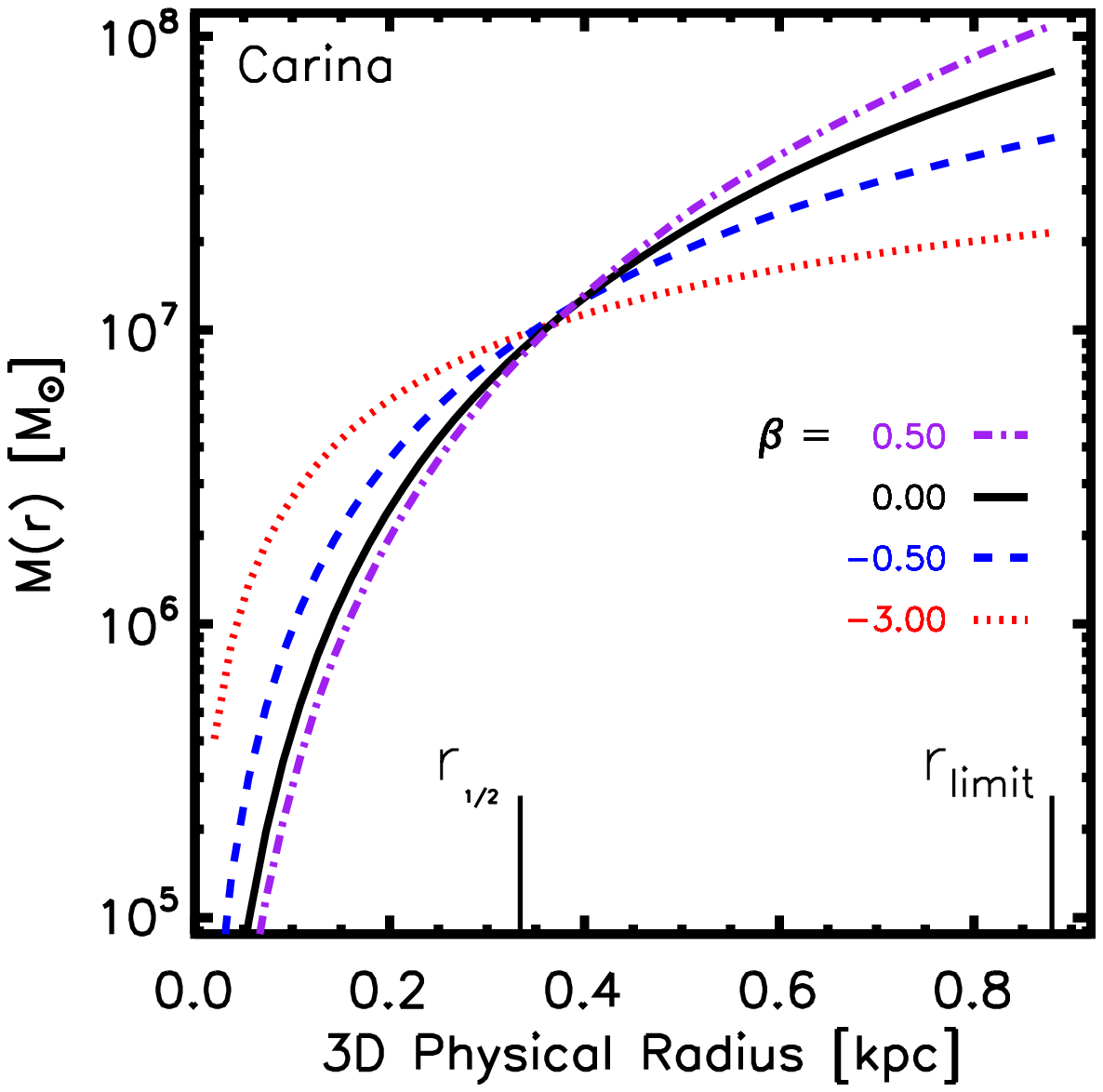}
\hspace{0.1mm}
\includegraphics[width=0.48\linewidth]{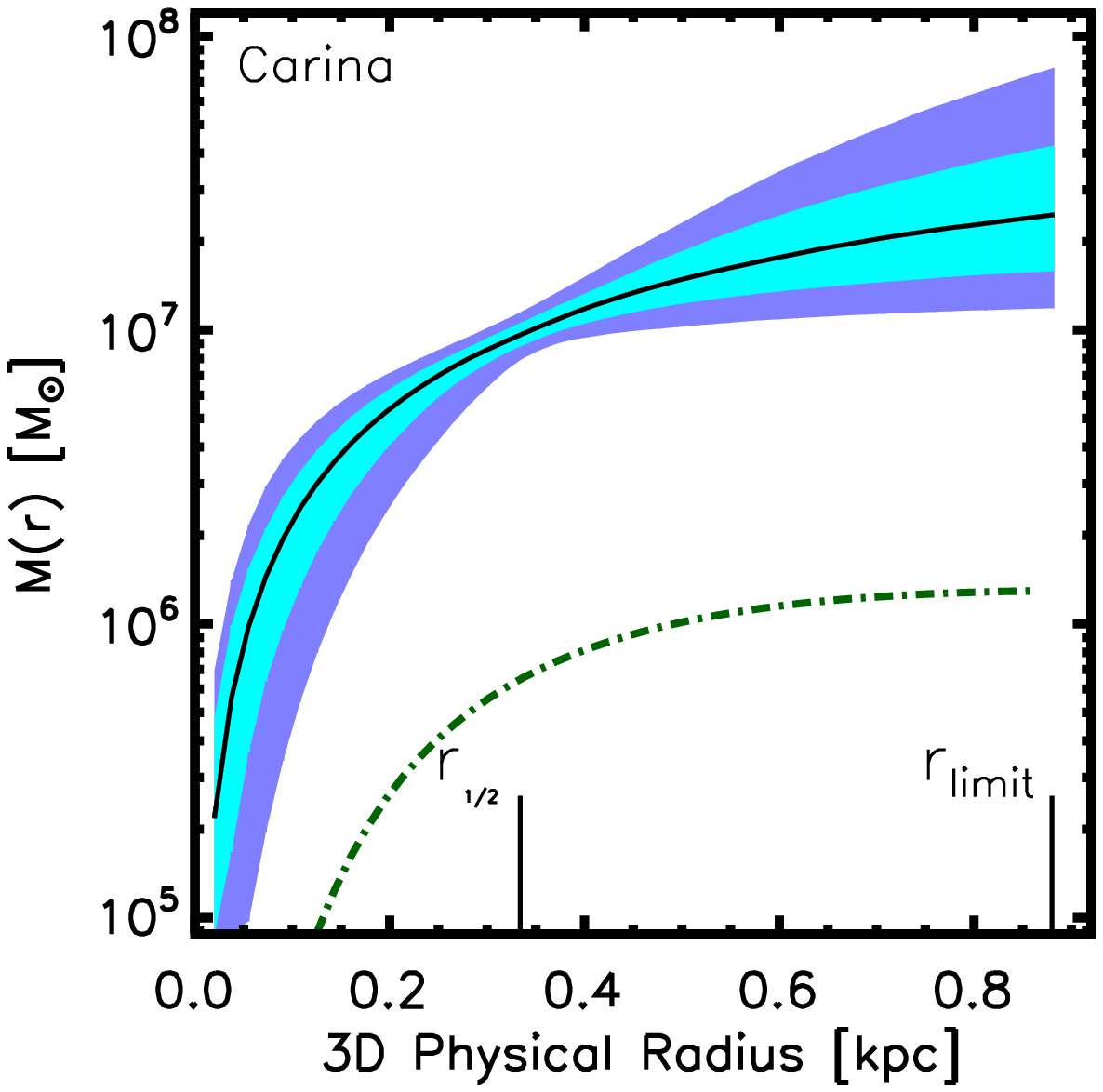}
\caption{{\em Left:} The cumulative mass profile generated by analyzing the Carina dSph using four different constant velocity dispersion anisotropies. The lines represent the median cumulative mass value from the likelihood as a function of physical radius. {\em Right:} The cumulative mass profile of the same galaxy, where the black line represents the median mass from our full mass likelihood (which allows for a radially varying anisotropy). The different shades represent the inner two confidence intervals (68\% and 95\%). The green dot-dashed line represents the contribution of mass from the stars, assuming a stellar V-band mass-to-light ratio of 3 $\Msun / \Lsun$.  This figure is from Wolf et al. (2010).
}
\label{fig:multibeta}
\end{figure*}

 It turns out that there is a single radius where the degeneracy between stellar velocity dispersion anisotropy and mass is minimized.
 As shown analytically and numerically by Wolf et al. (2010), this radius is close to the 3d half light radius, $\rhalf$, for most galaxies, and it is the mass with this radius, $\Mhalf = M(<\rhalf)$, that is best constrained by kinematic data.  Figure \ref{fig:multibeta} demonstrates this
 by showing explicitly mass model constraints for the Carina galaxy.
 By exploring models with variable $\beta$ values, several authors
 had seen similar results prior to the Wolf et al. (2010) derivation, but 
 with more restrictive conditions and without a full exploration of the implications (van der Marel 2000; Strigari et al. 2007a; Walker et al. 2009b).
Indeed, rather than extrapolate mass profiles to estimate a $\Vmax$ value for each dwarf halo,  Strigari et al. 2007b and 2008  advocated for the use of the integrated mass within a pre-defined radius that was similar in magnitude to the median $\rhalf$ value for the dwarf {\em population} (either 600 pc for large dwarfs, or 300 pc for smaller dwarfs) as a means of comparing data to simulations.  I will discuss these comparisons below, but first let me turn to a brief
 explanation of why the mass is best constrained within $\rhalf$.
 
 Qualitatively, one might expect that the degeneracy between the integrated mass and the assumed anisotropy parameter will be minimized
at some intermediate radius within the stellar distribution.  Think about the line-of-sight velocity dispersion measured  along the projected center ($R=0$) and then at the far edge $R = r_{\rm edge}$ of a spherical, dispersion supported galaxy that has stars extending to a 3d radius of $r_{\rm edge}$.  At the center,  line-of-sight observations will project onto the radial component with $\sigma_{\rm los} \sim \sigma_r$, while at the edge of the same galaxy, line-of-sight velocities project onto the tangential component with $\sigmalos \sim \sigma_t$.  Consider a galaxy that is intrinsically isotropic ($\beta = 0$). If this system is analyzed using line-of-sight velocities under the false assumption that $\sigma_r > \sigma_t$ ($\beta >0$), then the total velocity dispersion at $r \sim 0$ would be underestimated while the total velocity dispersion at $r \sim r_{\rm edge}$ would be overestimated. Conversely, if one were to analyze the same galaxy under the assumption that $\sigma_r < \sigma_t$ ($\beta < 0$), then the total velocity dispersion would be overestimated near the center and underestimated near the galaxy edge. It is plausible then that there is some intermediate radius where attempting to infer the enclosed mass from only line-of-sight kinematics is minimally affected by the unknown value of $\beta$.

\begin{figure*}[!t]
\centering
\begin{tabular}{ccc}
\includegraphics[width=4in]{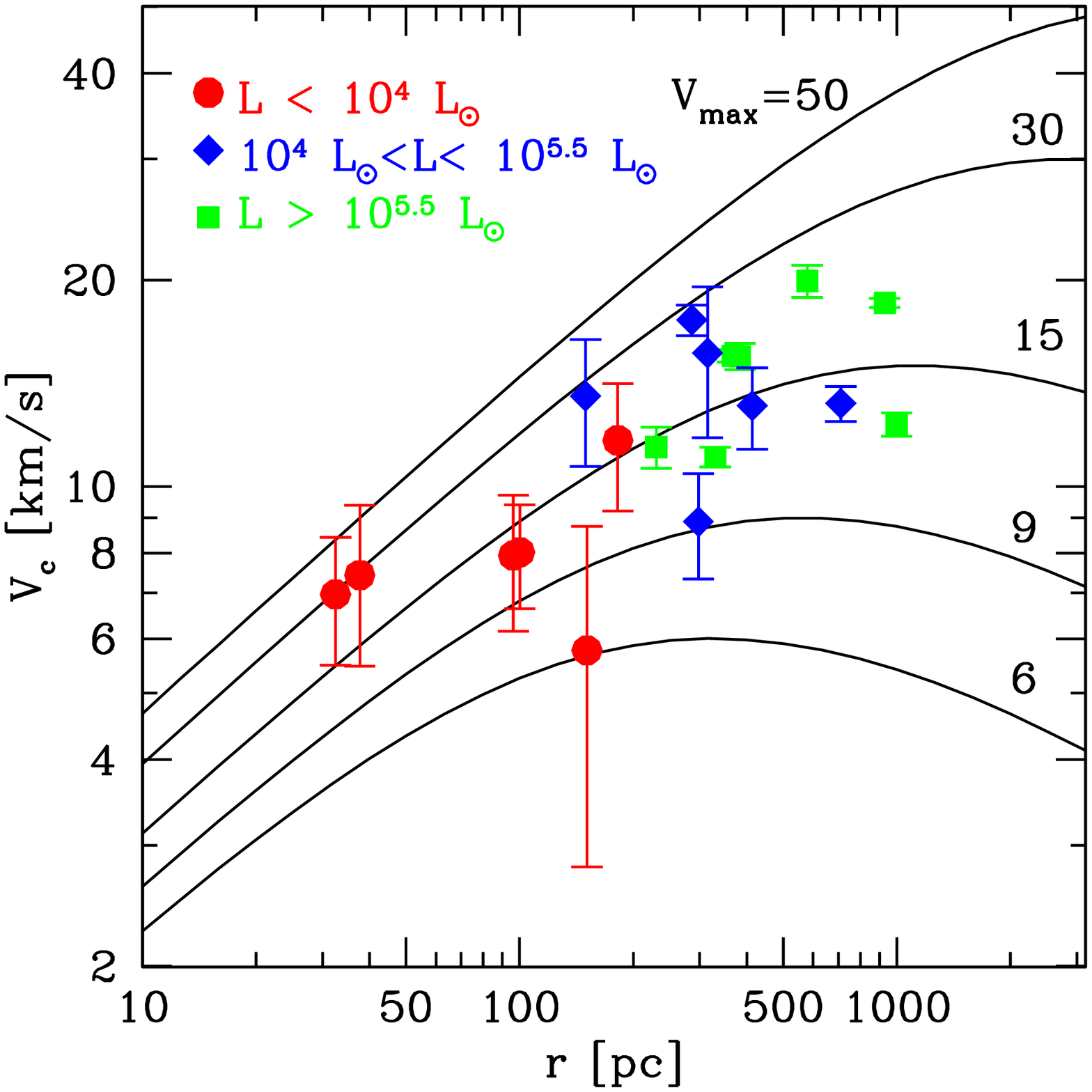}
\end{tabular}
\caption{Observed circular velocities $V_c(\rhalf)$ plotted versus $\rhalf$ for each of the Milky Way dSph galaxies discussed in Wolf et al. (2010).  The circular velocity curve values for the data were determined using Equation \ref{eq:vc}.  For reference, we plot $V(r)$ rotation curves for NFW subhalos obeying a {\em median} $\Vmax - \rmax$ relationship given by Equation \ref{eq:vm}.  Each curve is labeled by its $\Vmax$ value (assumed to be in $\kms$).  Dwarf galaxy points are color-coded by their luminosities (see legend).  Notice that the least luminous dwarfs (red) seem to fall along similar $V(r)$ curves as the most luminous dwarfs (green).}
\label{f:vvsr}
\end{figure*}

A more quantitative understanding of the $\rhalf$ mass constraint 
(but less rigorous than that provided in Wolf et al. 2010) can be gained by rewriting the Jeans equation such that
the $\beta(r)$ dependence is absorbed into the definition of 
$\sigmatot^2 = (3 - 2 \beta)\sigma_r^2$:
\begin{equation}
\label{eq:massjeans2}
G M(r) r^{-1} = \sigmatot^2(r) +  \sigma_r^2(r) \left (\gamma_\star+\gamma_\sigma -3 \right).
\end{equation}
It turns out that to very good approximation, the log slope of the stellar light profile at $\rhalf$ for a wide range of commonly-used stellar profiles is 
close to $-3$, such that  $\gamma_\star (\rhalf) = 3$ (Wolf et al. 2010).   Under the assumption that the velocity dispersion profile is flat 
(implying $\gamma_\sigma \ll 3$) 
we see that at $r=\rhalf$ the mass depends only on $\sigmatot$:
\begin{eqnarray}
M(\rhalf) & \simeq & G^{-1}\sigmatot^2(\rhalf) \, \rhalf \simeq 3 \, G^{-1} \sigmastar^2 \, \rhalf \,. 
\end{eqnarray}
 In the above chain of
arguments I have used the fact that $\sigmastar = \ave{\sigmatot^2}$ and the approximation $\ave{\sigmatot^2} \simeq \sigmatot^2(\rhalf)$.  This second
approximation works because the stellar-weighted velocity dispersion 
gets its primary contribution from the radius where $\gamma_\star = 3$ (Wolf et al. 2010).   A cleaner way to write this mass estimator is
\begin{equation}
V_c(\rhalf) = \sqrt{3} \, \sigmastar \, .
\label{eq:vc}
\end{equation}
Since by definition $\Vmax \ge V_c(\rhalf)$ we can conclude that 
\begin{equation}
\Vmax \ge \sqrt{3} \, \sigmastar .
\end{equation}
As mentioned above, a popular assumption in the literature has been to assume that dwarf galaxy subhalos obey $\Vmax = \sqrt{3} \, \sigmastar$  
(Klypin et al. 1999; Bullock et al. 2000; Simon \& Geha 2007).  It is likely that this common assumption under-estimated the $\Vmax$ values of MW and M31 satellite galaxies significantly.  Figure \ref{f:vvsr} plots $V_c(\rhalf)$ vs. $\rhalf$ for the sample of Milky Way dwarf galaxies discussed in Wolf et al. (2010).
For comparison, the lines show $V(r)$ rotation curves for NFW subhalos (Navarro et al. 1997) obeying the $\Vmax - \rmax$ relationship given by Equation \ref{eq:vm}.  Each curve is labeled by its $\Vmax$ value (assumed to be in $\kms$).  In many cases, the rotation curves continue to rise well beyond the $\rhalf$ value associated with each point.  Note that all of these galaxies are consistent with sitting in halos larger than $\Vmax = 10 \, \kms$, but that in many cases the implied extrapolation is significant.

\begin{figure*}[!t]
\centering
\begin{tabular}{ccc}
\includegraphics[width=4in]{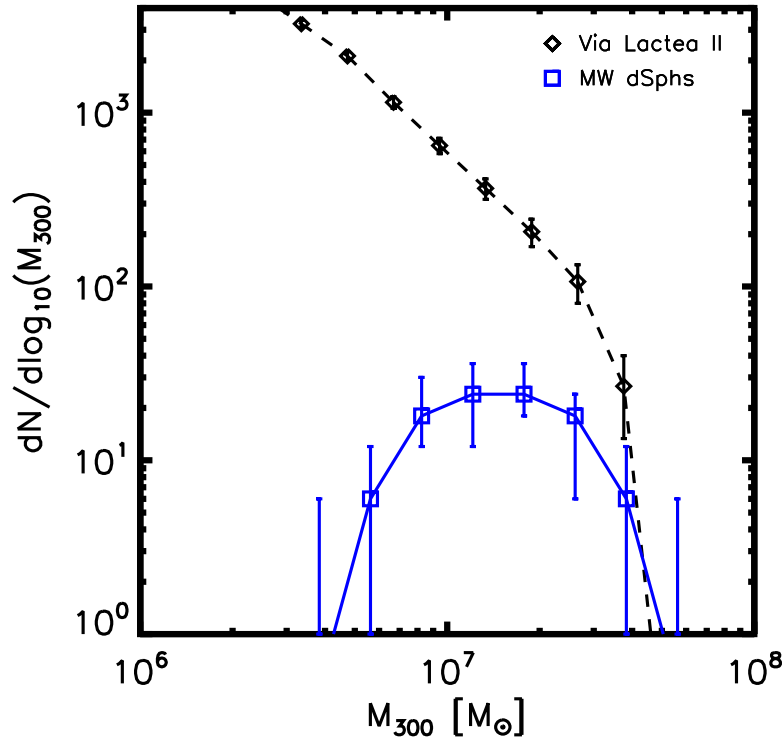}
\end{tabular}
\caption{Mass function for $M_{300} = M(< 300 {\rm pc})$ for MW dSph satellites and dark subhalos in the Via Lactea II  simulation within a radius of 400 kpc. The short-dashed curve is the subhalo mass function from the simulation. The solid curve is the median of the observed satellite mass function. The error bars on the observed mass function represent the upper and lower limits on the number of configurations that occur with a 98\% of the time (from Wolf et al., in preparation). Note that the mismatch is about $\sim 1$ order of magnitude at $M_{300} \simeq 10^7 \, \Msun$, and that it grows significantly towards lower masses.}
\label{f:m300mf}
\end{figure*}

\subsection{Mass within a Fixed Radius: $M_{600}$ and  $M_{300}$}

Strigari et al. (2007b) suggested that a more direct way to compare satellite galaxy kinematics to predicted subhalo counts was to consider their integrated masses within a radius that is as close as  possible to the median stellar radius for the galaxy population.  By focusing on the mass within a particular (inner) radius, one is less reliant on cosmology-dependent assumptions about how the dark matter halo density profiles are extrapolated to $\Vmax$.  This is particularly important if one is interested in simultaneously testing CDM models with variable $\sigma_8$ or if one is interested in more extreme variant models (e.g. dark matter from decays, Peter \& Benson 2010; or warm dark matter, Polisensky \& Ricotti 2010).    At the time when Strigari et al. (2007b) made their suggestion,  the smallest radius that was well resolved for subhalos in numerical simulations was $r \simeq 600$ pc (Via Lactea I; Diemand et al. 2007).  This motivated a comparison using $M_{600} = M (r< 600 \, {\rm pc})$, which is reasonably well constrained for the 9  `classical' (pre-SDSS) dSph satellites of the Milky Way.   They found that all but one of the classical Milky Way dSphs (Sextans) had an $M_{600}$ mass consistent with $(3-5) \times 10^{7} \Msun$.  By comparing to results from the Via Lactea I simulation, Strigari et al. (2007b) concluded that the Milky Way dwarf masses were
indicative of those expected if only the most massive halos (prior to accretion) were able to form stars.  

An updated version of this comparison (from Wolf et al., in preparation) is shown in Figure \ref{f:m300mf}, where now we focus on integrated masses within
300 pc.   This shift towards a smaller characteristic radius is motivated by an advance in simulation resolution, which now enables masses to be measured fairly accurately within 300 pc of subhalo centers.  It is fortuitous that 300 pc is also close to the median $\rhalf$ for Milky Way dSph galaxies with well-studied kinematic samples (Wolf et al. 2010; Bullock et al. 2010).   The short dashed line in Figure \ref{f:m300mf} shows the $M_{300}$ mass function from
the Via Lactea II simulation (Diemand et al. 2008), while the solid line shows $M_{300}$ for the known Milky Way dSph galaxies.  The masses are again indicative of a situation where only the most massive subhalos host galaxies.  Even at the point where the dSph mass function peaks ($M_{300} \simeq 10^7 \, \Msun$) the simulation over predicts the count by about a factor of 10.  This order of magnitude mismatch between observed counts and predicted counts {\em at fixed mass} is a reasonably conservative statement of  the MSP circa 2010.

\begin{figure*}[!t]
\centering
\begin{tabular}{ccc}
\includegraphics[width=4in]{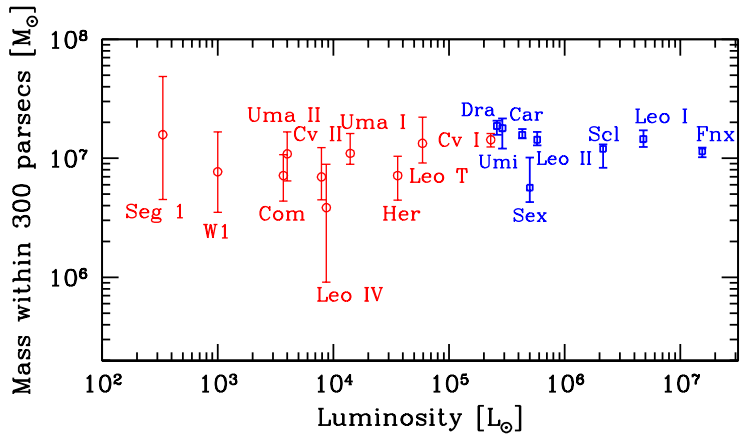}
\end{tabular}
\caption{Mass within 300 pc vs. V-band luminosity for classical (pre-SDSS) Milky Way dwarf satellites (blue) and ultra-faint (post-SDSS) satellites (red).
The most important point to take away from this plot is that there is {\em no obvious trend between mass and luminosity}.  
The trend seems to demonstrate scatter, but the masses do not appear to vary systematically with luminosity.}
\label{f:strigari}
\end{figure*}

The fact that the $M_{300}$ mass function of Milky Way satellites peaks sharply (within a factor of $\sim 4$) at a 
mass of $M_{300} \simeq 10^7 \, \Msun$ is remarkable 
given that these galaxies span a factor of $\sim 10^5$ in luminosity.  This point was highlighted by Strigari et al. (2008), who found that the $M_{300}$
mass-luminosity
relation for observed dwarfs is remarkably flat, with $M_{300} \propto L^{0.03 \pm 0.03}$.  The associated plot from Strigari et al. (2008) is shown 
in Figure \ref{f:strigari}.    The important point to take away from this plot is that there is {\em no detectable trend between galaxy mass and luminosity}. 
Galaxies with $L_V \simeq 4000 \, \Lsun$  (like Ursa Major II) demonstrate median $M_{300}$ masses that are very close to those of galaxies at  $L_V  \simeq 10^7 L_\odot$ (like Fornax).   Note that the Milky Way dwarfs do {\em not} all share {\em exactly} the same mass.  For example, Sextans is clearly less massive than Carina, Leo II, and Draco.  The mass of Hercules has been revised downward since the time that the plot in Figure \ref{f:strigari} was published (Ad{\'e}n et al. 2009) but this revision does not change the fact that there is a very weak relationship between mass and luminosity for the Milky Way dwarfs.

\begin{figure*}[!t]
\centering
\begin{tabular}{ccc}
\includegraphics[width=4in]{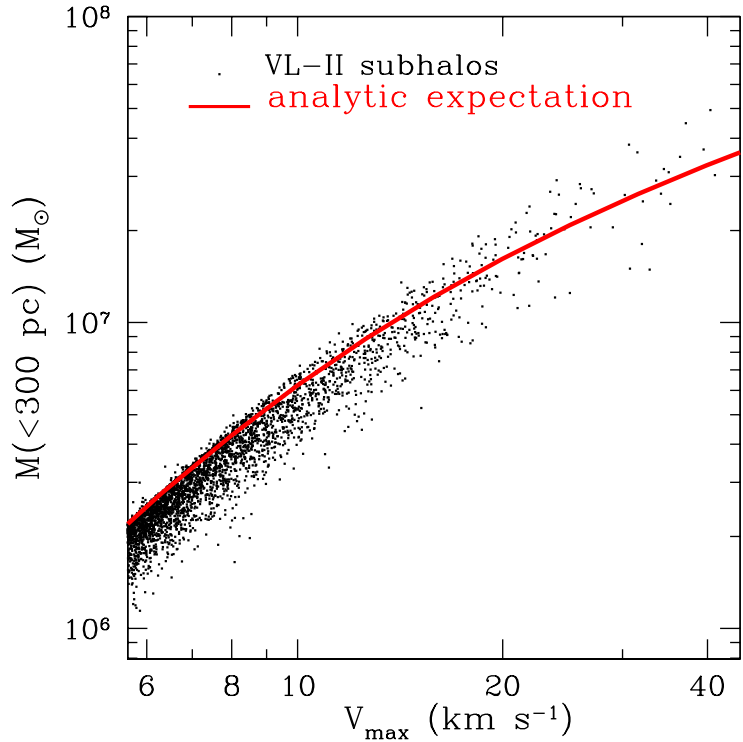}
\end{tabular}
\caption{Relationship between the mass within 300 pc and $\Vmax$ for subhalos in the Via Lactea II simulation (points from Diemand et al. 2008) along with the
analytic relationship expected for NFW halos that obey equation \ref{eq:vm} (red solid line).  }
\label{f:m300vmax}
\end{figure*}

Segue 1 is arguably the most interesting case.  With a luminosity of just $L_V \simeq 340 \, \Lsun$, the mass density of Segue 1 within its half-light radius ($\rhalf \simeq 38$ pc) is the highest of any Local Group
dwarf:  $\rho_{1/2} \simeq 1.6 \Msun$ pc$^{-3}$ (Martinez et al. 2010).  As can be seen by examining Figure \ref{f:vvsr} (red point, second from the left), 
a density this high at $\sim 40$ pc can only be achieved within a subhalo that is quite massive, with $\Vmax > 10 \, \kms$ or with $M_{300} \gtrsim 5 \times 10^{6} \Msun$.   Though the placement of Segue 1 on the Strigari plot requires significant extrapolation (from 40 pc to 300 pc), such an extrapolation is not unwarranted within the $\Lambda$CDM context because subhalos that are this dense and this massive almost always extend to tidal radii larger than 300 pc (as can be inferred, for example, from Equation \ref{eq:vm}).  With a luminosity some five orders of magnitude smaller than that of Fornax, Segue 1 is consistent with inhabiting a dark matter subhalo with approximately the same potential well depth.

Given that the $M_{300}$ mass variable is adopted for practical (not physical) reasons, it is worth examining its relationship to the more familiar measure $\Vmax$.  The relationship between the two variables is illustrated in Figure \ref{f:m300vmax}.  The points are Via Lactea II subhalos (kindly provided by the public release of Diemand et al. 2008) and the solid line is the analytic estimate for NFW subhalos that obey the $\Vmax - \rmax$ relation given by Equation \ref{eq:vm}.  
For $60 \, \kms \gtrsim \Vmax \gtrsim 15 \, \kms$, the mass within 300 pc correlates linearly with $\Vmax$ as $M_{300} \simeq 10^{7} \, \Msun (\Vmax/12.5 \, \kms)$.  
For smaller halos, the relationship asymptotes to $M_{300} \propto \Vmax^2$.  The point to take away from Figure \ref{f:m300vmax} is that the $M_{300}$ variable is at least as sensitive to potential well depth as the more familiar variable $\Vmax$.

\subsection{Making Sense of the Luminosity-Mass Relation}

\begin{figure*}[!t]
\centering
\begin{tabular}{ccc}
\includegraphics[width=4in]{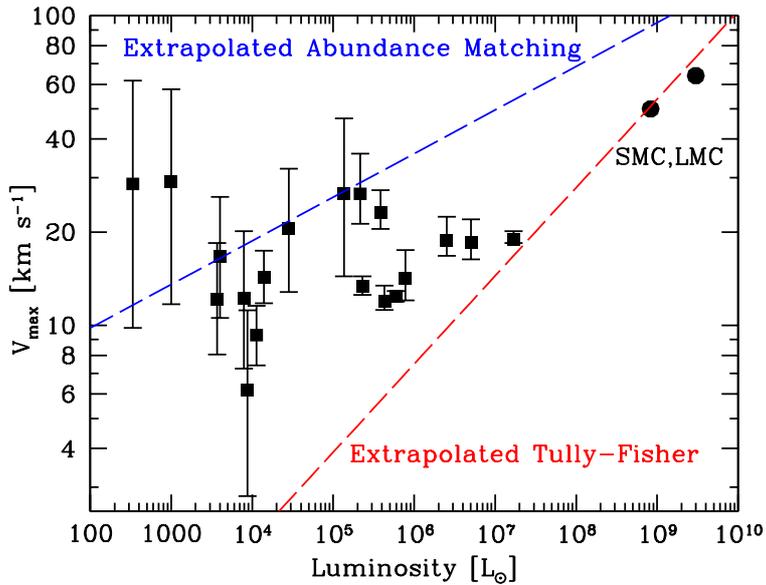}
\end{tabular}
\caption{The $\Vmax$ vs. V-band luminosity relation for the Milky Way satellite population, as inferred from assuming that dSph galaxies sit within NFW dark matter halos that obey the same scaling relations as do subhalos in $\Lambda$CDM N-body simulations.  The lower red dashed line is the Tully Fisher relation from Courteau et al. (2007) extrapolated to low lumnosities and the upper blue dashed line is the relation one obtains from extrapolating the abundance matching power-law from Busha et al. (2010).   }
\label{f:vmaxL}
\end{figure*}

 How does the observed relationship between luminosity and mass compare to simple empirical expectations that have been gained from examining more massive halos?  The symbols in Figure \ref{f:vmaxL} show inferred $\Vmax$ values for Milky Way dwarfs plotted as a function of luminosity.  I have opted to use $\Vmax$ as the scaling variable here (rather than $M_{300}$)  in order to make direct contact with more traditional scaling relations.  The error bars on each point are  typically larger (in a relative sense) than they are in Figure \ref{f:strigari} because the extrapolation to $\Vmax$ is more uncertain than the extrapolation to $M_{300}$ in most cases.  Nevertheless, the same global trend holds: there is a very weak correlation between luminosity and $\Vmax$ for the Milky Way dSph population.  Note that I have extended the luminosity axis slightly in order to include $\Vmax$ estimates for the LMC and SMC (from van den Bergh 2000, who discusses 
 the signifiant but unquantified uncertainties on mass estimates for the LMC and SMC).  The lower red dashed line is the extrapolated Tully Fisher relation~\footnote{The dSph satellites of the Milky Way also deviate from the Baryonic Tully Fisher relation (McGaugh \& Wolf 2010) as would be expected for systems that contain no detectable gas (except for Leo T; Grcevich \& Putman 2009).} for brighter spiral galaxies from Courteau et al. (2007)  and the upper blue dashed line is the relation one obtains from extrapolating the abundance matching power-law discussed  in reference to Figure \ref{f:am} above (specifically the $\Vmax - L$ relation advocated by Busha et al. 2010).  The fact that the naive extrapolation of the abundance matching power law provides a reasonable match at $L \simeq 10^4 \, \Lsun$ is encouraging.  Nevertheless, there are some surprising points of disagreement.

First,  the data clearly demonstrate a flatter trend with luminosity than either of these scaling relations.    The disagreement with the Tully Fisher line is less of a concern, as there is no reason to suspect that spheroidal galaxies should obey the same scaling relations as spiral galaxies.  Abundance matching, on the other hand, is a global measure of the $\Vmax - L$ mapping that is {\em required} to produce the correct abundances of galaxies.  Of particular interest in Figure \ref{f:vmaxL} is the mismatch between abundance matching and dwarf properties for the {\em most luminous dwarfs}.  The most luminous dSph galaxies at $L \sim 10^7 \, \Lsun$ prefer $\Vmax \simeq 20 \, \kms$, while the abundance matching expectation at this luminosity is much higher $\Vmax \simeq 50 \, \kms$. 
Note that a completely independent analysis by Strigari et al. (2010) find similarly low $\Vmax$ values for the luminous dSph galaxies.   Of course, one possible explanation is that the subhalos that host these objects  have lost a significant amount of dark matter mass, but several studies have looked at this effect in more detail and found that mass loss cannot easily account for this difference (Busha et al. 2010; Bullock et al. 2010).   Indeed, accounting for the fact that dSph satellite galaxies are subhalos simply lowers the normalization of the blue-dashed line in Figure \ref{f:vmaxL} by about 50 \%, without changing the slope to any appreciable agree.  The reason for this is that significant mass loss tends to have occurred only in systems that were accreted at early times.  At earlier times, larger $\Vmax$ values are required to produce the same stellar luminosity (e.g. Moster et al. 2010; Behroozi et al. 2010), and this compensates for the fact that $\Vmax$ tends to decrease with time once a subhalo is accreted.

One possible explanation for the lack of trend between $\Vmax$ (or $M_{300}$) and luminosity is that we are seeing evidence for a real scale in galaxy formation at $\Vmax \simeq 15 \, \kms$ or $M_{300} \simeq 10^7 \, \Msun$, which is remarkably close to the scale imposed by photoionization suppression and the HI cooling limit (Li et al. 2009; Macci{\`o} et al. 2009; Okamoto et al. 2009; Stringer et al. 2010).   This characteristic $\Vmax$/$M_{300}$ scale is well above the scale that would be indicative of a halo that required H$_2$ to cool: these are unlikely to be pre-reionization fossils of first light star formation 
(Madau et al. 2008; Ricotti 2010).   

A second explanation, which seems less likely, is that we are seeing a scale in dark matter clustering, which just happens to be very close to the mass scale where natural astrophysical suppression should kick in.  

A third possibility, addressed in the next section, is that 
the lack of an observed trend between mass and luminosity is the product of selection bias: most ultrafaint galaxies do inhabit halos with $M_{300} \lesssim 10^7 \Msun$,  but they are too diffuse to have been discovered.  If so, this implies that searches for the lowest mass ÒfossilÓ galaxies left over from reionization may be hindered by surface brightness limits (Bovill \& Ricotti 2009).

Finally, it is possible that the apparent flatness in the $\Vmax$/$M_{300}$ - luminosity relationship for Milky Way dwarfs is simply an accident of small statistics or an artifact of misinterpretation of the data.  Fortunately, as we now discuss, there is good reason to believe that the overall count of dwarf galaxies will grow significantly over the next decade.  These discoveries should enable larger statistical samples of dwarfs with kinematically-derived masses and a more stringent investigation of the trends that appear to be present in the data at the current time.

\section{Empirical Evidence for Missing Satellites}

The Sloan Digital Sky Survey (SDSS; Adelman-McCarthy et al. 2007) has revolutionized our understanding of the Milky Way's environment.  In particular, 
searches in SDSS data have more than doubled the number of known Milky Way satellite galaxies over the last several years (Willman et al. 2005; Zucker et al. 2006; Grillmair 2006, 2009; Majewski et al. 2007; Belokurov et al. 2007, 2009; Martin et al. 2009), revealing a population of galaxies that were otherwise 
too faint to have been discovered (with $L \lesssim 10^5 \, \Lsun$).    In addition to providing fainter detections, the homogeneous form of the SDSS has allowed for a much better understanding of the statistics of detection (Koposov et al. 2008; Walsh et al. 2009). 
The Milky Way satellite census is incomplete for at least two reasons:  sky coverage and luminosity bias.  A third source of incompleteness comes from the inability to detect and verify dwarf galaxies with very low surface brightness.  We turn to this last point at the end of this section.

\subsection{Sky Coverage}

Resolved star searches have covered approximately $\sim 20 \%$ of the sky, so it is reasonable to expect that there is a factor of  $c_{\rm sky} \simeq 5$  more satellites fainter than $L = 10^5 \, \Lsun$, bringing to the total
count of Milky Way dwarf satellites to $N \simeq N_{\rm pre-SDSS} + c_{\rm sky} \, N_{\rm SDSS} \simeq 9 + 5 \times 12 \simeq 70$.  

Note that the estimate $c_{\rm sky} \simeq 5$ assumes that there is no systematic angular bias in the satellite distribution. If the satellite distribution is anisotropic on the sky, the value of $c_{\rm sky}$ could be significantly different from 5.  If, for example, the SDSS happens to be viewing a particularly underdense region of the halo, then the correction for covering the whole sky could be quite large $c_{\rm sky} \gg 5$.  
Tollerud et al. (2008) used subhalos in Via Lactea I simulation (Diemand et al. 2007) 
to estimate the variance in $c_{\rm sky}$ for SDSS-size pointings within the halo and found  that $c_{\rm sky}$ can vary from just $\sim 3.5$ up to $\sim 8.3$ depending on the mock surveyÕs orientation.  In principle, then, one might expect as many as 110 satellites (or as few as 50) from the sky coverage correction alone.

\subsection{Luminosity Bias}

The second source of number count incompleteness, luminosity bias, is more difficult to quantify because it depends sensitively on the (unknown) radial distribution of all satellites.  Koposov et al. (2008) and Walsh et al. (2009) both found that detection thresholds are mostly governed by the distance to the object and the object luminosity, as long as the
surface density of the dwarf was brighter than about $\mu = 30$ mag arcsec$^{-2}$.  

\begin{figure*}[!t]
\centering
\begin{tabular}{ccc}
\includegraphics[width=4in]{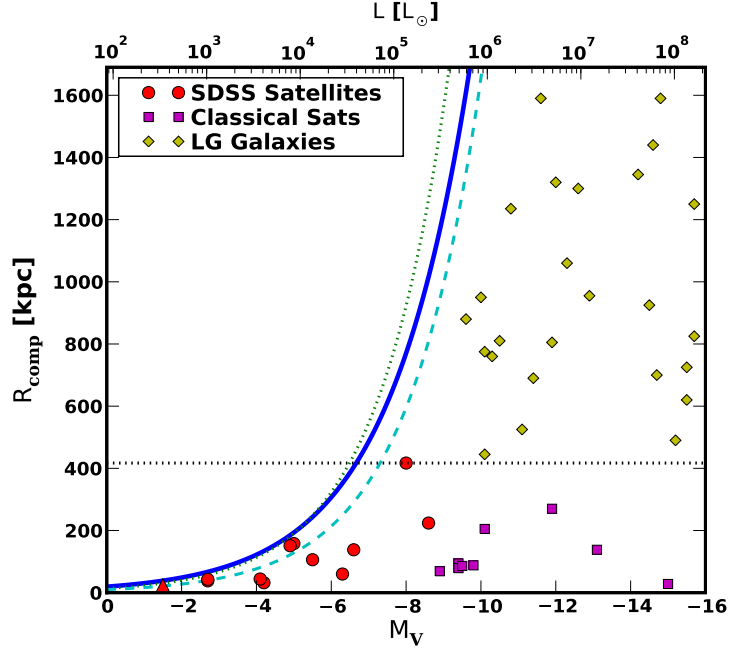}
\end{tabular}
\caption{The completeness radius for dwarf satellites from Tollerud et al. (2008). The three rising lines show the helio-centric distance, $R_{\rm{comp}}$, out to which dwarf satellites of a given absolute magnitude are complete within the SDSS DR5 survey.  The solid is for the published detection limits in Koposov et al. (2008) and the other lines are described in Tollerud et al. (2008).
 The dotted black line at 417 kpc corresponds to generous characterization of the outer edge of the Milky Way halo.  The data points are observed satellites of the Milky Way and the Local Group.  The red circles are the SDSS-detected satellites, the only galaxies for which the detection limits actually apply, although the detection limits nevertheless also delineate the detection zone for more distant Local Group  galaxies (yellow diamonds).  Purple squares indicate classical Milky Way satellites.  The faintest object (red triangle) is Segue 1, which is outside the DR5 footprint. }
\label{f:Rcomp}
\end{figure*}

Tollerud et al. (2008) used the Koposov et al. (2008) results to show that the SDSS is approximately complete down to a fixed apparent luminosity:
\begin{equation}
R_{\rm comp} \simeq 66 \, {\rm kpc} \left( \frac{L}{1000 \, \Lsun} \right)^{1/2} \, ,
\label{eq:Rcomp}
\end{equation}
where $R_{\rm comp}$ is a spherical completeness radius beyond which a dSph of a particular luminosity will go undetected.  
The implied relationship between galaxy luminosity and corresponding heliocentric completeness radius is shown by the solid curve in Figure \ref{f:Rcomp}.
The horizontal dotted line in Figure \ref{f:Rcomp} marks a generous estimate of $R_{\rm h} \simeq 417 \, {\rm kpc}$ for the Milky Way halo edge.
We see that only satellites brighter than $L \simeq 10^5 \, \Lsun$  are observable out to this radius. The fact that the faintest dwarf satellite galaxies known are more than 100 times fainter than this limit immediately suggests that there are many more faint satellite galaxies yet to be discovered.

\begin{figure*}[t!]
\centering
\begin{tabular}{ccc}
\includegraphics[width=4in]{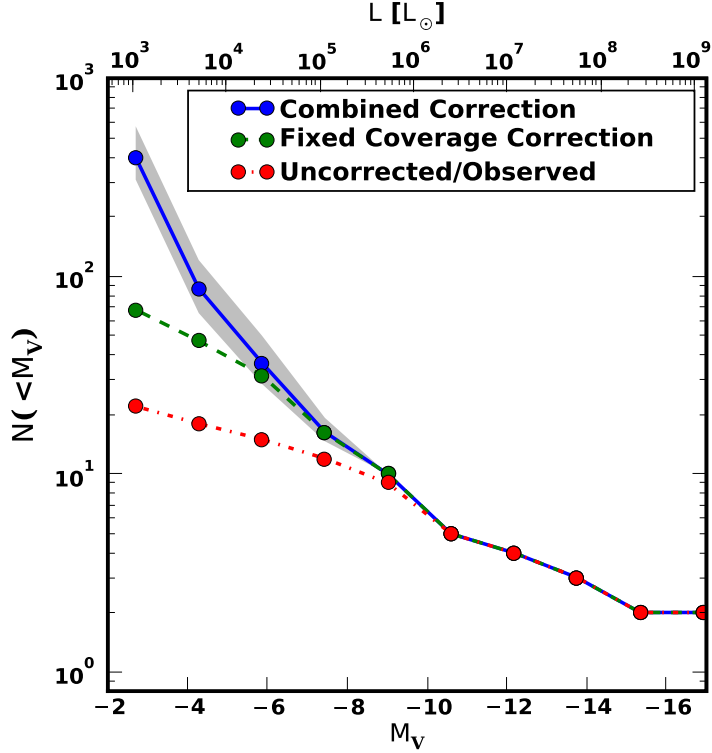}
\end{tabular}
\caption{Luminosity function of dSph galaxies within $R_{\rm h} = 417$ kpc of the Sun as observed (lower), corrected for only SDSS sky coverage (middle), and with luminosity completeness corrections from Tollerud et al. (2008) included (upper). Note that the brightest, classical (pre-SDSS) satellites are uncorrected, while new satellites have the correction applied.  The shaded error region corresponds to the 98\% spread over mock observation realizations within the Via Lactea I halo.}
\label{f:LF}
\end{figure*}

\begin{figure*}[t!]
\centering
\begin{tabular}{ccc}
\includegraphics[width=4in]{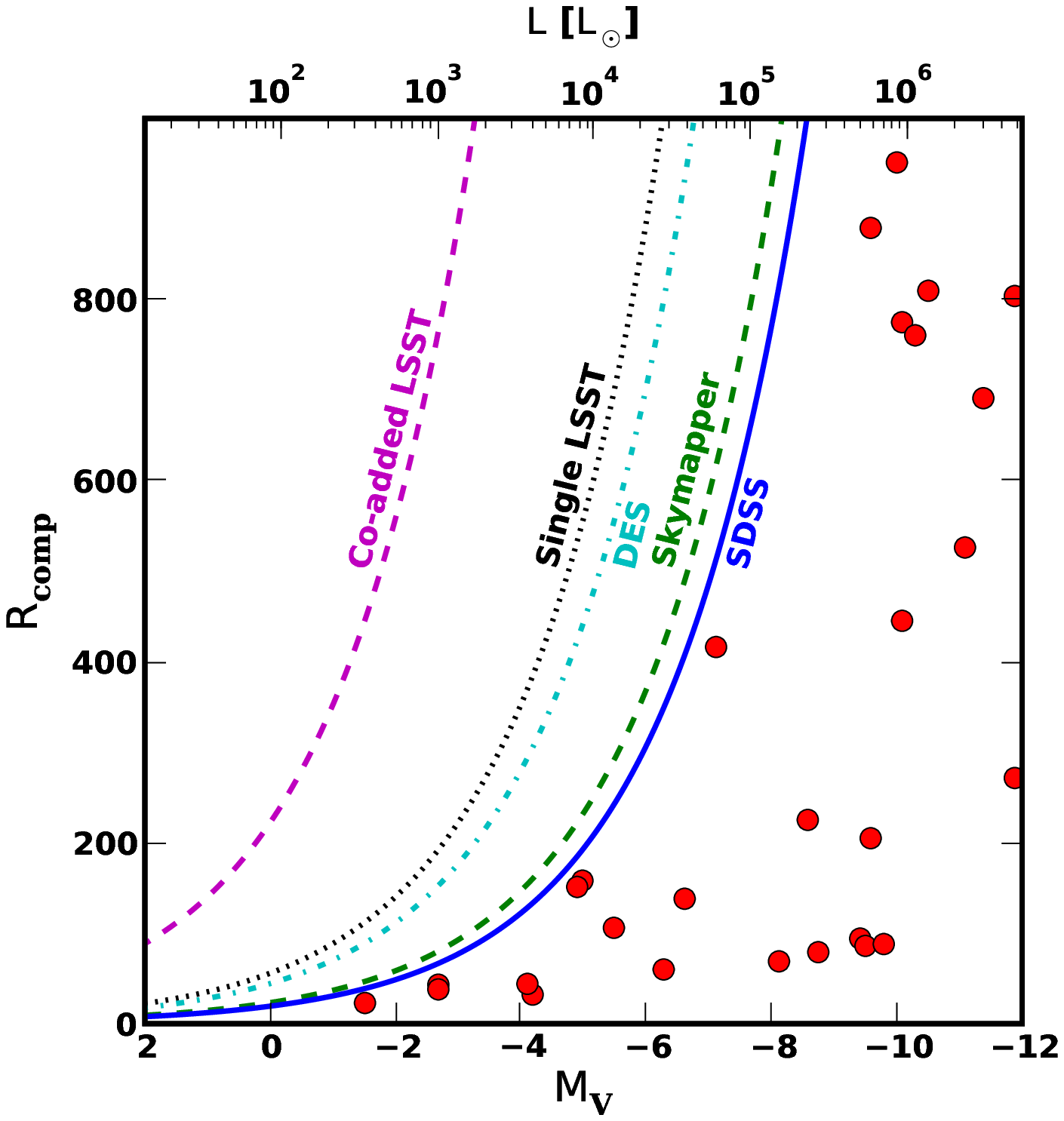}
\end{tabular}
\caption{Maximum radius for detection of dSphs as estimated by Tollerud et al. (2008) shown as a function of galaxy absolute magnitude for DR5 (assumed limiting r-band magnitude of 22.2) compared to a single exposure of LSST (24.5), co-added full LSST lifetime exposures (27.5), DES or one exposure from PanSTARRS (both 24), and the SkyMapper and associated Missing Satellites Survey (22.6). The data points are SDSS and classical satellites, as well as Local Group field galaxies.}
\label{f:Rcomp_future}
\end{figure*}

Once armed with a completeness radius - luminosity relation like that given in Equation \ref{eq:Rcomp}, one only needs to know the radial distribution of satellites $N(< R)$ in order to estimate the total count of faint galaxies out to some pre-defined edge of the Milky Way (we follow Tollerud et al. and use $R_{\rm h} = 417 \, {\rm kpc}$).    For example, if we knew that there were $N_{\rm obs}$ Milky Way dwafs  within $\Delta$ L of some  luminosity $L_{\rm obs}$ then  Equation \ref{eq:Rcomp} tells us that the census of these objects is complete to $R_{\rm comp}(L_{\rm obs})$.   We may estimate the total number of galaxies between $L_{\rm obs}$ and $L_{\rm obs} + \Delta L$ using 
\begin{equation}
N_{\rm tot} (L_{\rm obs}) \simeq c_{\rm sky} \, N_{\rm obs} \, \frac{N(<R_{\rm h})}{N(< R_{\rm comp}(L_{\rm obs}) )} \, .
\end{equation}
If we make the assumption that satellite galaxies are associated with subhalos in a one-to-one fashion, then $N(< R_{\rm h})/N(<R)$ may be estimated from analyzing the radial distribution of $\Lambda$CDM subhalos.  Tollerud et al. (2008) showed that the implied {\em ratio} $N(< R_{\rm h})/N(<R)$ is almost independent of how the subhalos are chosen.
As an example, consider the correction implied for the $N_{\rm obs} = 2$ known Milky Way dwarfs that have $L_{\rm obs} \simeq 1000 \Lsun$.
 For this luminosity, we are complete to $R_{\rm comp} = 66 \, {\rm kpc}$.  The subhalo distributions presented in Tollerud et al. (2008) obey  $N(<417 {\rm kpc})/N(< 66 {\rm kpc}) = 5 - 10$ for a wide range of subhalo population choices. This suggests a total count of $L \sim 1000 \, \Lsun$  galaxies is $N_{\rm tot}  \simeq 50 - 100$.

Figure \ref{f:LF} presents a more exacting correction of this kind from Tollerud et al. (2008).  The lower (red) curve shows the observed cumulative luminosity function of Milky Way dSph galaxies.  The middle (green) curve is corrected for SDSS sky coverage only and yields $\sim 70$ galaxies brighter than
$L = 1000 \Lsun$.     The upper curve has been corrected for
luminosity completeness (using Via Lactea I subhalos for N(<R)).  The result is that we expect  $\sim 400$ galaxies brighter than $1000 \, \Lsun$ within $\sim 400$ kpc of the Sun.  The correction becomes more significant for lower luminosity systems because the completeness radius is correspondingly smaller.
Note that this luminosity bias correction would be even more significant if one took into account the fact that the presence of a central disk will tend to deplete subhalos (and associated galaxies) in central regions of the Galactic halo (D'Onghia et al. 2010).   This effect will tend to increase the fraction of galaxy subhalos at large Galactocentric distances.

Referring back to  Figure \ref{f:Extrap} and Equation \ref{eq:nofv}, we see that $N \simeq 400$ satellites would be about the satellite count expected only for $\Vmax > 10 \kms$ halos.  It is encouraging that this is close to the $\Vmax$ limit inferred directly from stellar kinematics, as illustrated in
Figure \ref{f:vmaxL}.    Of course, as discussed in Tollerud et al. (2008) and Walsh et al. (2009), the overall luminosity bias correction is sensitive at the factor of 
$\sim 2$ level to the precise subset of subhalos that are used to determine the radial distribution of galaxies.  Generally, the best estimates suggest that there are hundreds of Milky Way satellite galaxies lurking in the outer reaches of the Milky Way.

There is real hope that these missing satellites will be detected as surveys like LSST, DES, PanSTARRS, and SkyMapper cover more sky and provide deeper maps of the Galactic environment (see Ivezic et al. 2008; The Dark Energy Survey Collaboration 2005; Kaiser et al. 2002; Keller et al. 2007).  Figure \ref{f:Rcomp_future} from Tollerud et al. (2008) provides a rough determination of the completeness radius for several planned surveys.  According to this estimate,  LSST will be able to detect objects as faint as  $L \sim 500 \, \Lsun$ out to the edge of the Milky Way halo.

\subsection{Surface Brightness Limits and Stealth Galaxies}
\label{s:stealth}

\begin{figure*}[t!]
\centering
\begin{tabular}{ccc}
\includegraphics[width=4in]{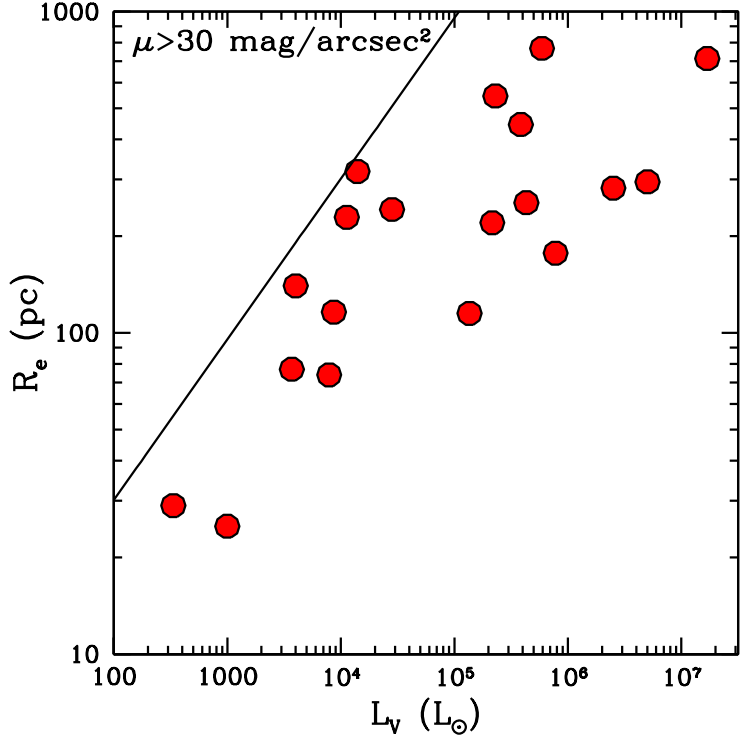}
\end{tabular}
\caption{Half-light radius $R_e$ vs. V-band luminosity for Milky Way dSph galaxies.
Galaxies above the solid line, with surface brightness fainter than $\mu = 30$
mag arcsec$^{-2}$ are currently undetectable.}
\label{f:sb}
\end{figure*}

 Importantly, the luminosity-distance detection limits discussed above only apply for systems with peak surface brightness obeying $\mu < 30$ mag arcsec$^{Ð2}$ (Koposov et al. 2008). Any satellite galaxy with a luminosity of  $L \simeq 1000 \Lsun$ and a  half-light radius $\rhalf$ larger than about 100 pc would have evaded detection with current star-count techniques regardless of its distance from the Sun.  This phenomenon is illustrated in Figure \ref{f:sb} where
 I plot the 2d (projected) half-light radius $R_{\rm e}$ vs. $L$ for Milky Way dSph galaxies.   The solid line shows a constant peak (central) surface brightness
of $\mu = 30$ mag arcsec$^{-2}$.   The tendency for many of the fainter dwarfs to line up near the surface brightness detection limit is suggestive. There is nothing ruling out the presence of a larger population of more extended systems that remain undetected because of their low surface brightness.  I refer to these undetected diffuse galaxies as {\em stealth galaxies}.

\begin{figure*}[t!]
\centering
\begin{tabular}{ccc}
\includegraphics[width=4in]{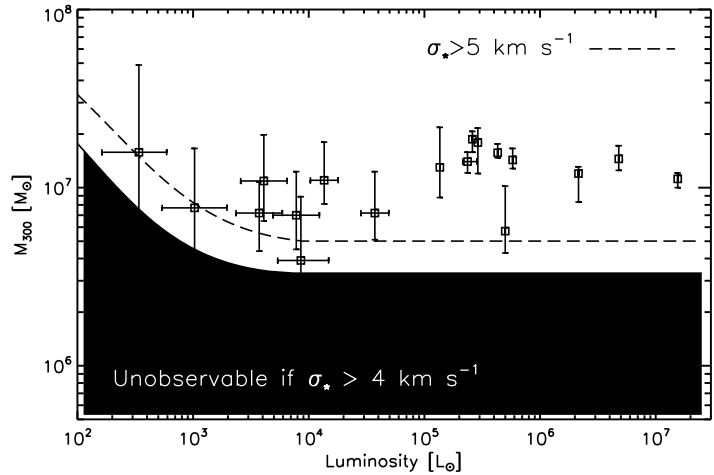}
\end{tabular}
\caption{Illustration of possible selection bias in the mass-luminosity plane of Milky Way dwarfs.  The data points with errors reproduce the Strigari et al. (2008) $M_{300}$ masses presented in Figure \ref{f:strigari}.   Galaxies within in the shaded region (below dashed line) will remain undetected ($\mu > 30$ mag arcsec$^{-2}$)     if they have velocity dispersions $\sigma_\star > \, 4 \kms$ ( $\sigma_\star > 5 \, \kms$). }
\label{f:stealth}
\end{figure*}

If a large number of stealth galaxies do exist, they are likely associated with low-mass dark matter subhalos.  If so, this will make it
 difficult to detect the lowest mass halos and will introduce a systematic bias that avoids low $M_{300}$ masses in the Strigari plot (Figure \ref{f:strigari}).

One can understand this expectation by considering a galaxy with velocity dispersion $\sigma_\star$ and luminosity $L$
embedded within a gravitationally-dominant dark matter halo described by a
circular velocity curve that increases with radius as an approximate power law:
$V_c(r) = V_{300} \, (r/300 \, {\rm pc})^\alpha$. 
Equation \ref{eq:vc} implies 
\begin{equation}
\rhalf = 300 \, {\rm pc} \, (\sqrt{3} \sigma_\star/ V_{300})^{1/\alpha} \, .
\end{equation}
For an NFW halo (Navarro et al. 1997) with scale radius $\rmax \gg 600~ {\rm pc}$ we have $\alpha = 1/2$ and 
\begin{equation}
 \rhalf \propto \left( \frac{\sigma_\star}{V_{300}}\right)^{2} \propto \frac{\sigma_\star^2}{M_{300}}  \, .
 \end{equation}
Clearly, $\rhalf$ increases  as we decrease $M_{300}$ at fixed $\sigma_\star$.
One implication is that if a galaxy has a stellar surface density $\Sigma_\star \propto L/\rhalf^2$ that is just large enough to be detected,  
another galaxy with identical $L$ and $\sigma_\star$ will be undetectable if it happens to reside within a slightly less massive halo.

Figure \ref{f:stealth} from Bullock et al. (2010) provides a more explicit demonstration of how surface brightness bias can effect the mass-luminosity relation of
dwarf galaxies.  The points show  $M_{300}$ vs. $L$ for Milky Way dSph galaxies, with masses from Strigari et al. (2008) and luminosities
updated as in Wolf et al. (2010).  The region below the shaded band is undetectable for dwarf galaxies with $\sigma_\star > \, 4 \, \kms$.
The region below the dashed line is undetectable for dwarf galaxies with  $\sigma_\star > \, 5 \, \kms$.   

This result suggests that surface brightness selection bias may play a role explaining the lack of observed correlation between luminosity and mass
  for Milky Way satellites.
   It also implies that searches for the lowest mass `fossil' galaxies left over from reionization may be hindered by surface brightness limits.  This latter point was
 made earlier by  Bovill \& Ricotti (2009).     According to estimates in Bullock et al. (2010), potentially half of several hundred satellite galaxies that could be observable by surveys like LSST are stealth. 
  A complete census of these objects  will require deeper sky surveys,
30m-class follow-up telescopes, and more refined methods to identify
extended, self-bound  groupings of stars in the halo.

\section{Summary}

Advances in simulation technology have solidified the decade-old expectation that substructure should be abundant in and around CDM halos, with counts that rise steadily to the smallest masses.  Properties of observed satellites in and around the Local Group provide an important means to test this prediction.

  The substructure issue has gained relevance over the years because it touches a range of important issues that span many subfields, including the microphysical nature dark matter, the role of feedback in galaxy formation, and star formation in the early universe.  It is in fact difficult to state directly what it would take to solve the MSP without first associating with a specific subfield.  From a galaxy formation standpoint, one important question is to identify the primary feedback sources that suppress galaxy formation in small halos.  An associated goal is to identify any obvious mass scale where the truncation in the efficiency of galaxy formation occurs.

It is particularly encouraging for $\Lambda$CDM theory that both direct kinematic constraints on the masses of Milky Way satellite galaxies and completeness correction estimates both point to about the same mass and velocity scale, $\Vmax \simeq 15 \, \kms$ or $M_{300} \simeq 10^7 \, \Msun$.   The luminosity bias correction discussed in association with Figure \ref{f:LF} suggests that the Milky Way halo hosts about 400 satellite galaxies with luminosities similar to the faintest dwarf galaxies known.   As seen in either Figure \ref{f:Extrap} or Figure  \ref{f:Vmaxfunc}, a total count of 400 satellites is
approximately what is expected at a minimum $\Vmax$ threshold of $15 \, \kms$ -- the same $\Vmax$ scale evidenced in the kinematic measurements shown in Figure \ref{f:vmaxL}.  These numbers did not have to agree.  Interestingly, they both point to a mass scale that is close to limit where cooling via atomic hydrogen is suppressed and where photoionization heating should prevent the accretion of fresh gas (see Dekel 2005).

More puzzling is the overall lack of observed correlation between Milky Way satellite galaxy luminosities and their $M_{300}$ masses or 
$\Vmax$ values (see Figures \ref{f:strigari} and \ref{f:vmaxL}).  Most of the models that have been constructed to confront the mass-luminosity
over-predict $\Vmax$ values at for the brightest dwarfs and under-predict them for the faintest dwarfs (Li et al. 2009; Macci{\`o} et al. 2009; Okamoto et al. 2009; Busha et al. 2009; Bullock et al. 2010).  These differences may reflect small-number statistics at the bright end, and a potential observational bias associated with surface brightness at the faint end.

Section \ref{s:stealth} discussed the idea that there may be population of low-luminosity satellite galaxies orbiting within the halo of the Milky Way that are too diffuse to have been detected with current star-count surveys, despite the fact that they have luminosities similar to those of known ultrafaint  dSphs.  
These stealth galaxies should preferentially inhabit the smallest dark matter subhalos that can host stars ($V_{\rm max} \lesssim 15 ~ \kms$).
  One implication is that selection bias may play a role explaining the lack of observed correlation between $L$ and $M_{300}$ for Milky Way satellite galaxies.
 This effect also implies that searches for the lowest mass `fossil' galaxies left over from reionization may be hindered by surface brightness limits.
 A large number (more than one hundred) stealth galaxies may be orbiting within the Milky Way halo.  These systems have stellar distributions too diffuse to have been easily discovered so far, but new surveys and new techniques are in the works that may very well reveal hundreds of these missing satellite galaxies 
within the next decade. 

\bigskip

\begin{center}
{\bf{Acknowledgments}}
\end{center}

I am most grateful to David Martinez Delgado for organizing this lively and enriching winter school and to the other lecturers and students who made the experience  enjoyable.     I would like to thank Peter Behroozi, Michael Kuhlen, and Joe Wolf for allowing me to include their unpublished figures here.  I would also like to acknowledge my recent collaborators on MSP work for influencing my thoughts on the subject:  Marla Geha, Manoj Kaplinghat, Greg Martinez, Quinn Minor, Josh Simon, Kyle Stewart, Louis Strigari, Erik Tollerud, Beth Willman, and Joe Wolf.    I thank Erik Tollerud and Joe Wolf for providing useful comments on the manuscript.

\end{document}